\documentclass[aps,pre,twocolumn,10pt,showpacs,amsfonts,amssymb,amsmath,balancelastpage,footinbib]{revtex4-1}
\usepackage{bm}
\usepackage{graphicx}
\usepackage{subfigure}
\usepackage[pdftex]{hyperref}
\usepackage{fancyhdr}
\usepackage{multirow}
\begin{document}
%
%
%
%
\title{Coexistence of phases and the observability of random graphs}
\author{Antoine \surname{Allard}}
\author{Laurent \surname{H\'ebert-Dufresne}}
\author{Jean-Gabriel \surname{Young}}
\author{Louis J. \surname{Dub\'e}}
\affiliation{D\'epartement de physique, de g\'enie physique et d'optique, Universit\'e Laval, Qu\'ebec (Qc), Canada G1V 0A6}
\date{\today}
\begin{abstract}
 In a recent Letter, Yang \textit{et al.} [Phys. Rev. Lett. \textbf{109}, 258701 (2012)] introduced the concept of \textit{observability transitions}: the percolation-like emergence of a macroscopic \textit{observable} component in graphs in which the state of a fraction of the nodes, and of their first neighbors, is monitored. We show how their concept of depth-$L$ percolation---where the state of nodes up to a distance $L$ of monitored nodes is known---can be mapped unto multitype random graphs, and use this mapping to exactly solve the observability problem for arbitrary $L$. We then demonstrate a non-trivial coexistence of an observable and of a non-observable extensive component. This coexistence suggests that monitoring a macroscopic portion of a graph does not prevent a macroscopic event to occur unbeknown to the observer. We also show that real complex systems behave quite differently with regard to observability depending on whether they are geographically-constrained or not.
\end{abstract}
\maketitle
%
%
%
\section{Introduction}
%
Considered as the ultimate proof of our understanding, the \textit{controllability} (and its dual concept the \textit{observability}) of natural and technological complex systems have been the subject of many recent studies \cite{Liu11_Nature,Saadatpour11_PLoSComputBiol,Sahasrabudhe11_NatCommun,Cowan12_PLoSONE,Liu12_PLoSONE,Nepusz12_NaturePhys,Yan12_PhysRevLett,Yang12_PhysRevLett,Liu13_ProcNatlAcadSciUSA,Posfai13_SciRep,Jia13_NatCommun,Sun13_PhysRevLett,Jia13_SciRep}. In essence, the question is whether the global state of a system can be imposed (inferred) through the control (monitoring) of a few of its constituents. By mapping the underlying web of interactions between the constituents of systems unto graphs, analytical criteria have been proposed to determine whether a system is controllable (observable), and if so, through which of its constituents control (monitoring) should be applied. However, although promising and theoretically correct, doubts have been raised as to whether these criteria can be used in practice on real large systems \cite{Cowan12_PLoSONE,Sun13_PhysRevLett}.

Whenever a comprehensive and exact theoretical framework is lacking, simpler but solvable theoretical models---that consider simplified versions of the systems under scrutiny---become valuable alternatives to highlight and understand key behaviors of complex systems. Following this trend, Yang \textit{et al.} used random graphs to study the observability of power grids through the use of \textit{phasor measurement units} that allow to monitor the state of a node and of each of its neighbors \cite{Yang12_PhysRevLett}. Using this approach, they demonstrated that the largest observable component emerges in a percolation-like transition, and argued that structural properties found in real systems reduce the number of monitoring units required for achieving large-scale observability.

We formalize their approach into the general concept of depth-$L$ percolation where the state of nodes up to a distance $L$ of monitored nodes is known as well. Using a multitype version of the Configuration Model \cite{Allard09_PhysRevE}, we study analytically the emergence of the extensive ``giant'' observable component (i.e., its size and the conditions for its existence), and we demonstrate a non-trivial coexistence with another extensive component: one made of non-monitored nodes. We then turn our attention to graphs extracted from real complex systems and show that many such systems support the coexistence of two extensive components. Moreover, our theoretical framework yields analytical arguments to explain the low thresholds for the large-scale observability observed in many of these systems. However, we find that geographically-constrained systems (e.g., power grids) are poorly modeled by random graphs; rather, their topology appears similar to the one of lattices. Our results also suggest that they behave quite differently with regard to observability: their structure does not support coexistence, and achieving large-scale observability requires more monitoring units than hinted by calculations based on the Configuration Model \cite{Yang12_PhysRevLett}.

This paper is organized as follows. In Sec.~\ref{sec:coex_depthL_perco}, we introduce the concept of depth-$L$ percolation and develop an exact mathematical description for the case $L=1$. This allows one to demonstrate the equivalence between our approach and the one proposed in Ref.~\cite{Yang12_PhysRevLett}, and to identify the possible coexistence of two extensive components. In Sec.~\ref{sec:coex_gen_math_from}, we generalize our mathematical framework to any $L$, and use it to study the effect of varying the depth on the coexistence regime. In Sec.~\ref{sec:coex_real}, we investigate the observability of graphs extracted from real complex systems with numerical simulations and our mathematical framework. Conclusions and final remarks are collected in the last section. A technical Appendix is supplied to describe the case $L=2$ and to compare it with the results obtained in Ref.~\cite{Yang12_PhysRevLett}.
\begin{figure}[t]
  \includegraphics[width=0.4\textwidth]{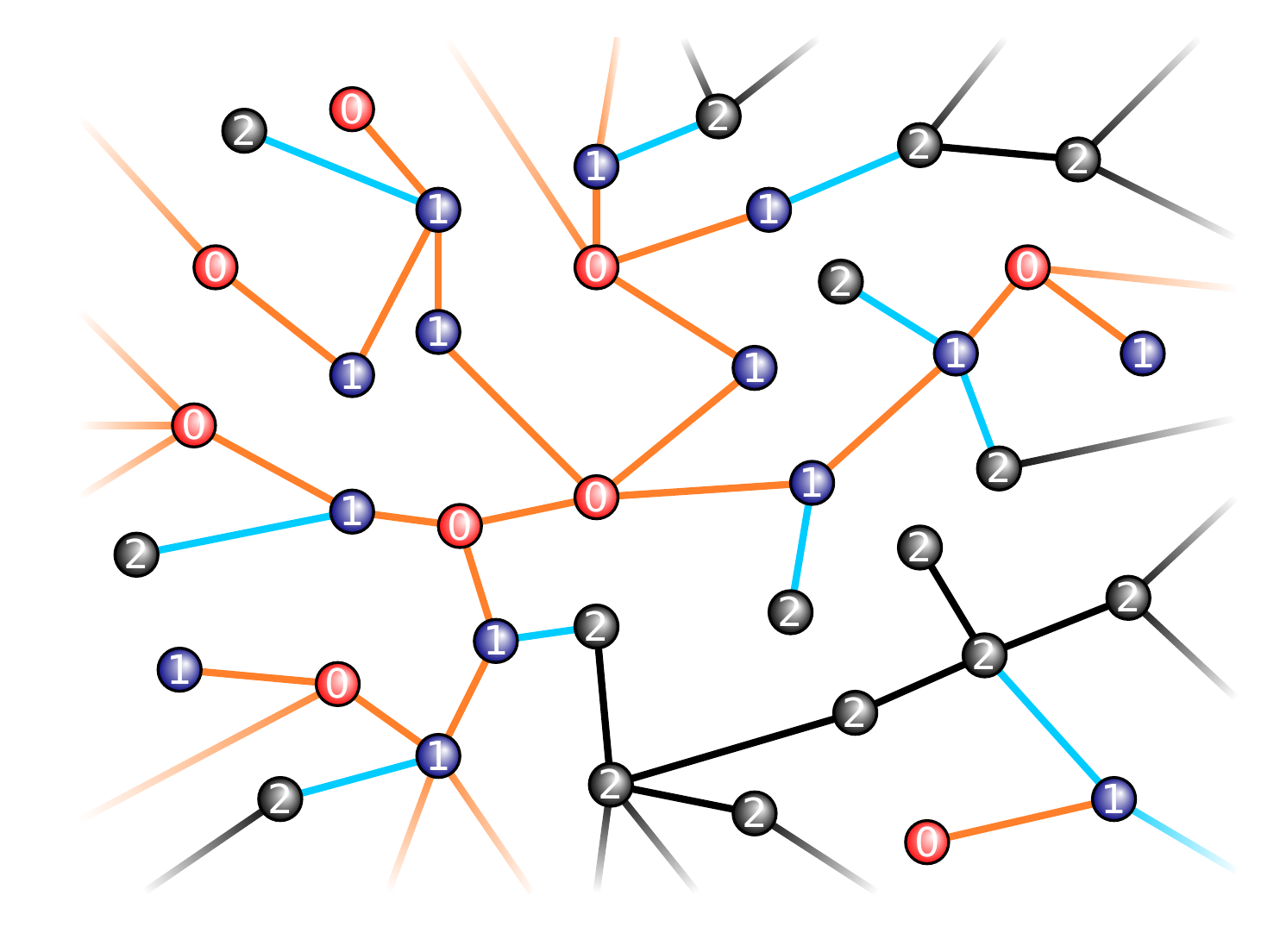}
  \caption{\label{fig:coex_example_L1}(color online). Illustration of depth-$1$ percolation on a graph generated through the Configuration Model. Directly occupied, indirectly occupied and non-occupied nodes are in red (type 0), blue (type 1) and black (type 2), respectively. Occupied components are identified with orange edges, and non-occupied components with black edges. Edges linking occupied and non-occupied components---the ones removed by setting $x_{02}=x_{12}=x_{20}=x_{21}=1$ in Eqs.~\eqref{eq:coex_g0},~\eqref{eq:coex_g1}~and~\eqref{eq:coex_g2}---are shown in cyan.}
\end{figure}
%
%
%
%
%
\section{Depth-$L$ percolation} \label{sec:coex_depthL_perco}
%
Depth-$L$ percolation is a generalization of traditional site percolation: nodes are occupied independently with probability $\varphi$ \textit{and} every node up to a distance $L$ of occupied nodes are also occupied. We say that the latter are \textit{indirectly occupied} as opposed to the former which are said to be \textit{directly occupied} (see Fig.~\ref{fig:coex_example_L1}). Depth-0 percolation corresponds to traditional site percolation (see Sec.~\ref{sec:coex_case_L0}). For the sake of simplicity (and to make an explicit correspondence with the mathematical treatment in Ref.~\cite{Yang12_PhysRevLett}), we first focus on depth-1 percolation---where first neighbors of occupied nodes are occupied as well---on graphs generated through the Configuration Model \cite{Newman10_Networks}. The generalization to any $L$ is however straightforward in our formalism and is the subject of Sec.~\ref{sec:coex_gen_math_from}.

The Configuration Model defines a maximally random graph ensemble whose graphs are random in all respects except for the degree distribution, $\{P(k)\}_{k\in\mathbb{N}}$, prescribing the number of connections that nodes have (i.e., number of first neighbors). Using probability generating functions (pgf), many exact results can be obtained in the limit of large graphs \cite{Newman01_PhysRevE,Newman02_PhysRevE,Allard12_JPhysA}. For the present study, we define the pgf associated with the degree distribution
\begin{align} \label{eq:coex_G0}
  G_0(x) & = \sum_{k=0}^{\infty} P(k) x^k \ ,
\end{align}
and the one generating the number of edges \textit{leaving} a node reached by one of its edges (\textit{excess} degree distribution)
\begin{align} \label{eq:coex_G1}
  G_1(x) & = \frac{G_0^{\prime}(x)}{G_0^{\prime}(1)} = \frac{1}{\langle k \rangle} \sum_{k=1}^{\infty} kP(k)x^{k-1} \ .
\end{align}
Here the prime denotes the derivative, and $\langle k \rangle$ corresponds to the first moment of the degree distribution (i.e., average degree). The Configuration Model generates graphs through a stub pairing scheme: a random number of stubs (the degree) is assigned to each node according to $\{P(k)\}_{k\in\mathbb{N}}$, and edges are formed by randomly matching stubs together. In the context of depth-$L$ percolation, directly occupied nodes are then selected with probability $\varphi$, and the identification of indirectly occupied nodes follows.
%
%
%
\subsection{Mapping to multitype random graphs}
%
To study the emergence of the extensive occupied component, we introduce a mapping linking depth-$L$ percolation to percolation on multitype random graphs \cite{Allard09_PhysRevE}. To facilitate this mapping, we consider an alternative procedure to generate graphs with directly and indirectly occupied nodes. As previously stated, a degree is assigned to each node according to $\{P(k)\}_{k\in\mathbb{N}}$, but instead of pairing stubs right away, directly occupied nodes (type 0) are first selected with probability $\varphi$. Stubs of type 0 nodes are then randomly matched with any stubs in the graph; untagged nodes now connected to type 0 nodes are said to be indirectly occupied (type 1). Note that two type 0 nodes can be linked together. Nodes that have neither been tagged as type 0 nor as type 1 are said to be non-occupied (type 2). All remaining free stubs are finally paired randomly to \textit{close} the graph. This alternative perspective is identical to the one discussed in the previous section in the limit of large graphs, and is analog to \textit{on-the-fly} network construction \cite{Noel12_PhysRevE}. Although it may seem unnecessary in the simple case $L=1$, this slight change of perspective greatly eases the generalization to an arbitrary value of $L$.

By definition, a fraction $w_0=\varphi$ of the nodes is of type 0. Because these nodes are assigned randomly and independently, the distribution of the number of connections they have with other node types (their \textit{joint} degree distribution) is
\begin{align} \label{eq:coex_P0}
  P_0(\bm{k}) & = \delta_{0k_2} P(k_0+k_1) \frac{(k_0+k_1)!}{k_0!k_1!} \varphi^{k_0} (1-\varphi)^{k_1} \ ,
\end{align}
where $\bm{k}=(k_0,k_1,k_2)$ and $\delta_{ab}$ is the Kronecker delta. In other words, if a neighbor of a type 0 node is not of type 0, it is inevitably of type 1. The associated pgf is
\begin{align}
  g_0(\bm{x}) & = \sum_{\bm{k}} P_0(\bm{k}) x_{00}^{k_0} x_{01}^{k_1} x_{02}^{k_2} \nonumber \\
              & = G_0\big(\varphi x_{00} + [1-\varphi]x_{01}\big) \ . \label{eq:coex_g0}
\end{align}
A randomly chosen node will be of type 1 if it has not been selected as a type 0 node and if at least one of its neighbors is of type 0. This happens with probability $(1-\varphi)[1-(1-\varphi)^k]$ for a node whose degree is equal to $k$. Averaging over the degree distribution, we find
\begin{subequations} \label{eq:coex_w}
 \begin{align}
  w_1 & = (1-\varphi) [1-G_0(1-\varphi)] \label{eq:coex_w1} \ .
 \end{align}
 Asking for normalization, we find that type 2 nodes represent a fraction
 \begin{align}
  w_2 & = (1-\varphi) G_0(1-\varphi) \label{eq:coex_w2}
 \end{align}
\end{subequations}
of the nodes. Likewise, we define $\varepsilon_i$ as the probability that a randomly chosen edge leads to a type $i$ node. Clearly $\varepsilon_0=\varphi$, and by similar arguments as above but by averaging over the excess degree distribution instead, we find that
\begin{subequations} \label{eq:coex_epsilon}
 \begin{align}
  \varepsilon_1 & = (1-\varphi) [1 - G_1(1-\varphi)] \label{eq:coex_epsilon1} \\ 
  \varepsilon_2 & = (1-\varphi) G_1(1-\varphi) \label{eq:coex_epsilon2} \ .
 \end{align}
\end{subequations}
From the alternative procedure described above, we find that the joint degree distribution of the union of type 1 and type 2 nodes is
\begin{align}
  P_{1\bigcup2}(\bm{k}) & = P(k_0\!+\!k_1\!+\!k_2) \frac{(k_0\!+\!k_1\!+\!k_2)!}{k_0!k_1!k_2!} \varepsilon_0^{k_0} \varepsilon_1^{k_1} \varepsilon_2^{k_2} \ .
\end{align}
Since the difference between nodes of these two types is the presence of type 0 nodes in their immediate neighborhood, we can readily write the pgf associated with their joint degree distribution
\begin{align}
  g_1(\bm{x}) & = A_1 \sum_{\bm{k}} (1-\delta_{0k_0}) P_{1\bigcup2}(\bm{k}) x_{10}^{k_0} x_{11}^{k_1} x_{12}^{k_2} \label{eq:coex_g1} \\
              & = \frac{G_0\big(\varepsilon_0 x_{10}\!+\!\varepsilon_1 x_{11}\!+\!\varepsilon_2 x_{12} \big)\!-\!G_0\big(\varepsilon_1 x_{11}\!+\!\varepsilon_2 x_{12} \big)}{1-G_0\big(1-\varphi\big)} \nonumber
\end{align}
and
\begin{align}
  g_2(\bm{x}) & = A_2 \sum_{\bm{k}} \delta_{0k_0} P_{1\bigcup2}(\bm{k}) x_{20}^{k_0} x_{21}^{k_1} x_{22}^{k_2} \nonumber \\
              & = \frac{G_0\big(\varepsilon_1 x_{21} + \varepsilon_2 x_{22} \big)}{G_0\big(1-\varphi\big)} \label{eq:coex_g2} \ ,
\end{align}
where $A_1$ and $A_2$ are normalization constants. With $\{w_i\}_{i=0,1,2}$ and the pgf $\{g_i(\bm{x})\}_{i=0,1,2}$ in hand, we are now in a position to mathematically describe the emergence of the giant occupied component.
%
%
%
\subsection{Giant occupied component}
\begin{figure}[t]
  \subfigure[\ $\langle k \rangle = 1.7$, $L=1$]{\label{fig:coex_validation_void_L1}    \includegraphics[width=0.44\textwidth]{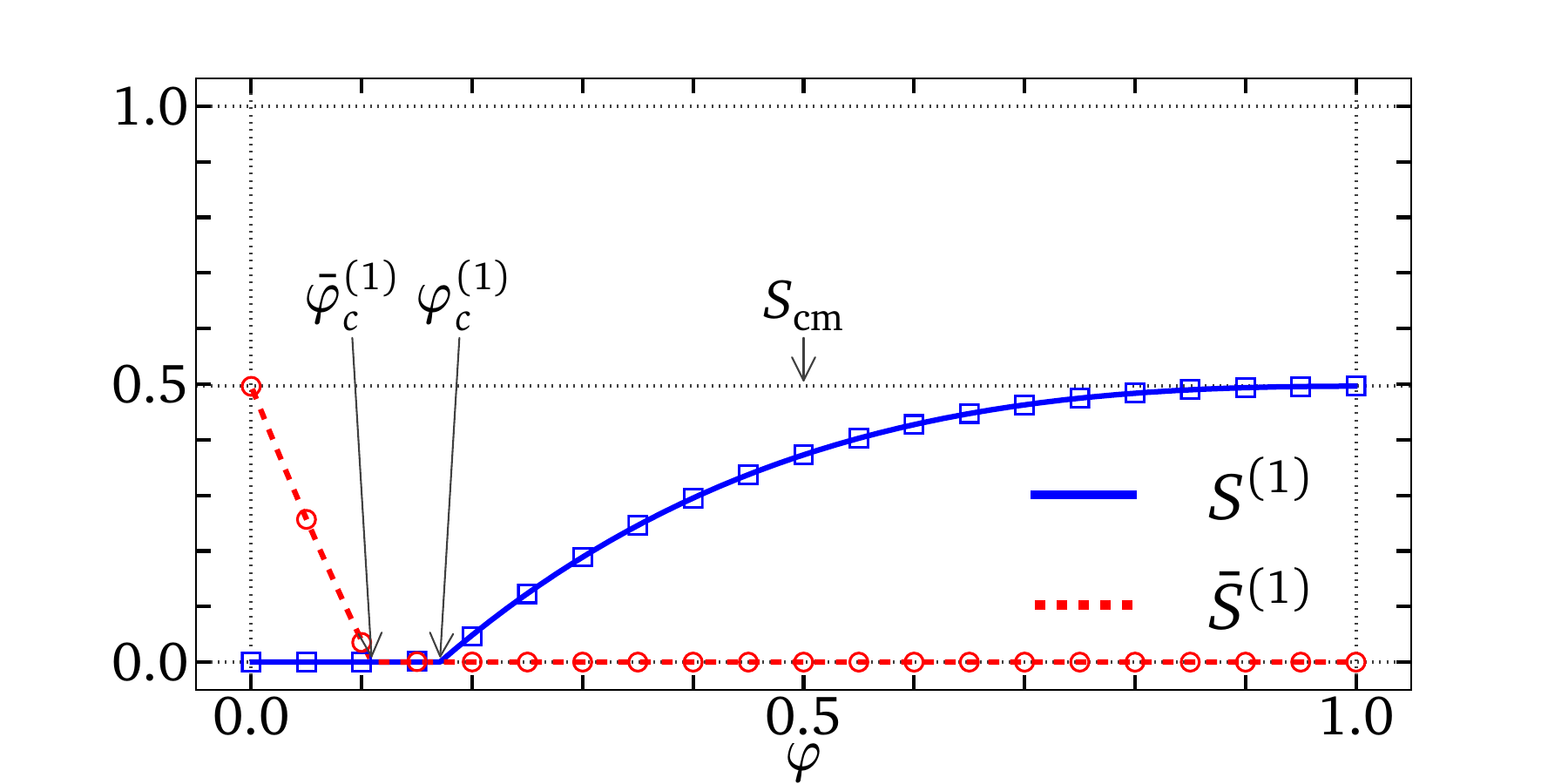}}\\
  \subfigure[\ $\langle k \rangle = 2.0$, $L=1$]{\label{fig:coex_validation_overlap_L1} \includegraphics[width=0.44\textwidth]{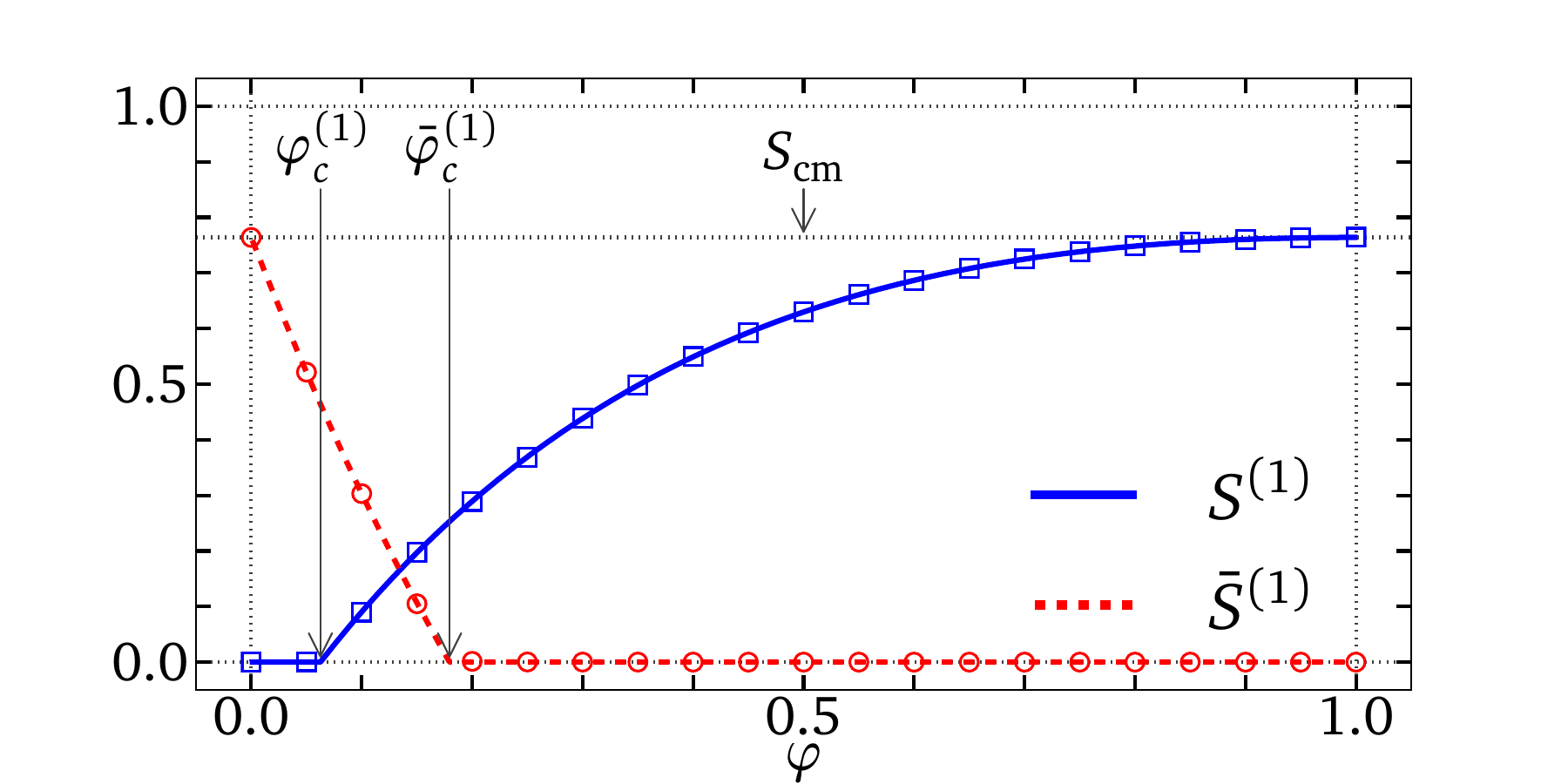}}\\
  \subfigure[\ $\langle k \rangle = 2.0$, $L=3$]{\label{fig:coex_validation_overlap_L3} \includegraphics[width=0.44\textwidth]{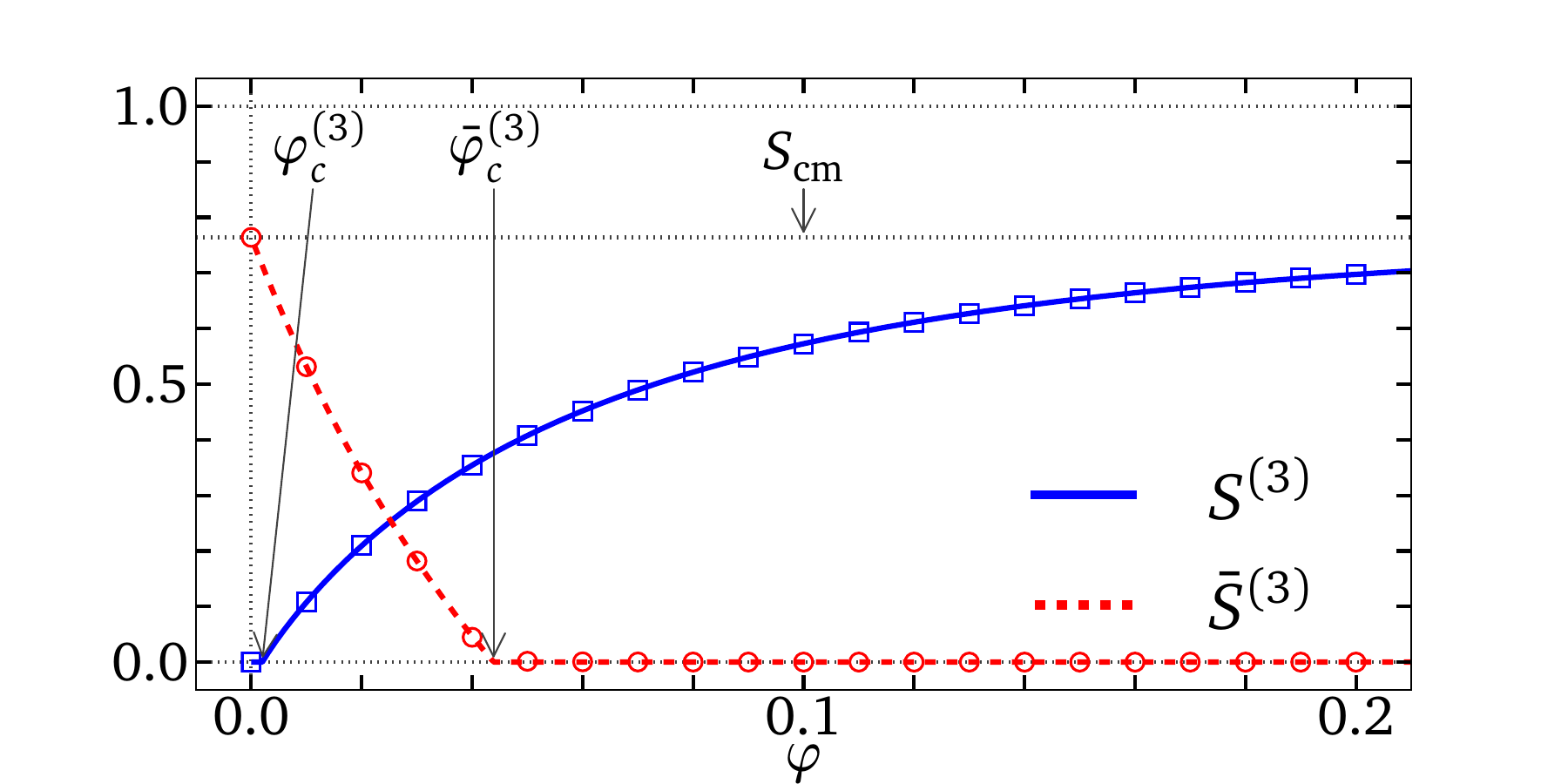}}
  \caption{\label{fig:coex_validation}(color online). Validation of the theoretical formalism for different depth of percolation ($L$). Size of the occupied and non-occupied components in function of $\varphi$ for graph ensembles with a different average degree. The degrees of both ensembles are exponentially distributed according to $P(k)=(1-\mathrm{e}^{-\lambda})\mathrm{e}^{-\lambda (k-1)}$ with $k \geq 1$ and $\lambda=-\ln\big( 1 - 1 / \langle k \rangle \big)$. The size of the giant component in the original graph ensemble, $S_\mathrm{cm}$, is shown for comparison. It is equal to $S_\mathrm{cm}=1-G_0(a)$ where $a$ is the solution of $a=G_1(a)$ \cite{Newman01_PhysRevE}. Curves are the solutions of Eqs.~\eqref{eq:coex_goc_a}--\eqref{eq:coex_goc_S} and \eqref{eq:coex_gnc_a}--\eqref{eq:coex_gnc_S} [$L=1$], and Eqs.~\eqref{eq:coex_gen_goc_S}--\eqref{eq:coex_gen_goc_a} and \eqref{eq:coex_gen_gnc_S}--\eqref{eq:coex_gen_gnc_a} [$L=3$]. Symbols are the relative size of the largest occupied and non-occupied components averaged over at least 100 graphs of at least $5\times10^{5}$ nodes each. Threshold values were obtained from Eqs.~\eqref{eq:coex_goc_threshold}, \eqref{eq:coex_gnc_threshold} and \eqref{eq:coex_gen_gnc_threshold}, and by analyzing the stability of Eqs.~\eqref{eq:coex_gen_goc_a} around $\bm{a}=\bm{1}$. Note the change of scale of the abscissa of \protect\subref{fig:coex_validation_overlap_L3}.}
\end{figure}

%
It has been shown in Ref.~\cite{Allard09_PhysRevE} that the relative size of the giant component, $\mathcal{S}$, in multitype random graphs is computed via
\begin{align} \label{eq:coex_mult_S}
  \mathcal{S} & = \sum_{i=0}^{M-1} w_i \big[1 - g_i(\bm{a}) \big]
\end{align}
where $M$ is the number of node types, and where $\bm{a}=\{a_{ij}\}_{i,j=0,\ldots,M-1}$ is the set of probabilities that an edge leaving a type $i$ node towards a type $j$ node \textit{does not} lead to the giant component. These probabilities correspond to the stable fixed point---the smallest solution in $[0,1]^{M^2}$---of the following system of equations
\begin{align} \label{eq:coex_mult_a}
  a_{ij} & = \frac{\partial g_{j}(\bm{a})/\partial x_{ji}}{\partial g_{j}(\bm{1})/\partial x_{ji}}
\end{align}
with $i,j=0,\ldots,M-1$. We are interested in the relative size of the giant occupied component, the component made of type 0 and type 1 nodes solely. To do so, we isolate them from type 2 nodes by setting $x_{02}=x_{12}=x_{20}=x_{21}=1$ in Eqs.~\eqref{eq:coex_g0},~\eqref{eq:coex_g1}~and~\eqref{eq:coex_g2}. Noting that $a_{10}=a_{00}$, this yields
\begin{subequations} \label{eq:coex_goc_a}
  \begin{align}
    a_{00} & = G_1\big(\varphi a_{00} + (1-\varphi) a_{01}\big) \label{eq:coex_goc_a00} \\
    a_{01} & = G_1\big( \varphi a_{00} + \varepsilon_1 a_{11} + \varepsilon_2 \big) \label{eq:coex_goc_a01} \\
    a_{11} & = \frac{G_1\big( \varphi a_{00} + \varepsilon_1 a_{11} + \varepsilon_2 \big) - G_1\big( \varepsilon_1 a_{11} + \varepsilon_2 \big)}{1 - G_1\big( 1-\varphi\big)} \label{eq:coex_goc_a11}
  \end{align}
\end{subequations}
and the relative size of the giant occupied component, $S$, becomes [summing Eq.~\eqref{eq:coex_mult_S} only over $i=$ 0, 1]
\begin{align} \label{eq:coex_goc_S}
  \begin{split}
    S = &\ 1 - \varphi G_0\big( \varphi a_{00} + (1-\varphi)a_{01} \big) \\
        & - (1-\varphi) \Big\{ G_0\big( \varphi a_{00} + \varepsilon_1 a_{11} + \varepsilon_2 \big) \\
        & + G_0\big(1-\varphi\big) - G_0\big( \varepsilon_1 a_{11} + \varepsilon_2 \big) \Big\} \ .
  \end{split}
\end{align}
Clearly, $a_{00}=a_{01}=a_{11}=1$ is always a solution of Eqs.~\eqref{eq:coex_goc_a} and corresponds to the situation where there is no giant occupied component ($S=0$). A giant occupied component emerges in fact when the stable fixed point $\bm{a}=\bm{1}$ undergoes a transcritical bifurcation during which a stable fixed point appears in $[0,1)^3$. Hence a linear stability analysis of $\bm{a}=\bm{1}$ leads to the criterion
\begin{align} \label{eq:coex_goc_threshold}
  \begin{split}
    G_1^{\prime}(1) = &\ 1 + (1-\varphi_c)G_1^{\prime}(1-\varphi_c)\\ & \times \biggl\{ 1 - \varphi_c G_1^{\prime}(1) - \varphi_c(1-\varphi_c)\bigl[G_1^{\prime}(1)\bigr]^2 \biggr\}
  \end{split}
\end{align}
marking the point $\varphi=\varphi_c$ where the giant occupied component starts to emerge. This is the exact same criterion obtained by Yang \textit{et al.} \cite{Yang12_PhysRevLett}. In fact, by identifying $u \equiv \varphi a_{00} + (1-\varphi)a_{01}$ and $(1-\varphi)s \equiv \varepsilon_1 a_{11} + \varepsilon_2$, Eqs.~\eqref{eq:coex_goc_a}--\eqref{eq:coex_goc_S} fall back on their results, thereby demonstrating the equivalence between the two approaches. Notice also that we retrieve from Eqs.~\eqref{eq:coex_goc_a}--\eqref{eq:coex_goc_threshold} the well-known results for the Configuration Model \cite{Newman01_PhysRevE} in the limit $\varphi\rightarrow1$ (see the caption of Fig.~\ref{fig:coex_validation}).

The multitype perspective offers a simple interpretation of the emergence of the giant occupied component. For such component to exist, the original graph ensemble---defined by $\{P(k)\}_{k\in\mathbb{N}}$---must itself have a giant component, which occurs when $G_1^{\prime}(1)>1$ \cite{Newman01_PhysRevE}. A giant \textit{occupied} component then exists if an extensive component composed of only type 0 and type 1 nodes prevails after edges between occupied and non-occupied nodes have been removed (cyan edges in Fig.~\ref{fig:coex_example_L1}). In other words, a giant occupied component exists if the original giant component is robust to the removal of these edges. Yet if the original giant component is \textit{robust enough}, an extensive component composed of non-occupied nodes solely could also prevail, therefore leading to the coexistence of two extensive components.
%
%
%
\subsection{Giant non-occupied component}
%
One equation has been left out of Eqs.~\eqref{eq:coex_goc_a}. Indeed, Eqs.~\eqref{eq:coex_mult_a} yields another nontrivial equation
\begin{align} \label{eq:coex_gnc_a}
  a_{22} & = \frac{G_1\big(\varepsilon_1 + \varepsilon_2 a_{22}\big)}{G_1\big(1-\varphi\big)}
\end{align}
for the probability that an edge between two type 2 nodes does not lead to an extensive component. Since this component is made of non-occupied nodes solely, we refer to it as the \textit{giant non-occupied component}. By summing Eq.~\eqref{eq:coex_mult_S} over type 2 nodes only, the relative size of this other extensive component is
\begin{align} \label{eq:coex_gnc_S}
  \bar{S} & = w_2 \left[ 1 - \frac{G_0\big( \varepsilon_1 + \varepsilon_2 a_{22} \big)}{G_0\big( 1-\varphi \big)} \right] \ .
\end{align}
Again, we see that $a_{22}=1$ is always a solution of Eq.~\eqref{eq:coex_gnc_a} and the point $\varphi=\bar{\varphi}_{c}$ at which it becomes an instable fixed point, that is when
\begin{align} \label{eq:coex_gnc_threshold}
  (1-\bar{\varphi}_{c})G_1^{\prime}\big(1-\bar{\varphi}_{c}\big) = 1 \ ,
\end{align}
marks the (dis)appearance of the giant non-occupied component. Again, notice that Eqs.~\eqref{eq:coex_gnc_a}--\eqref{eq:coex_gnc_threshold} fall back on the results for the Configuration Model \cite{Newman01_PhysRevE} in the limit $\varphi\rightarrow0$ (see the caption of Fig.~\ref{fig:coex_validation}).
%
%
%
\subsection{Coexistence of extensive components}
%
Figure~\ref{fig:coex_validation} depicts the typical scenarios with respect to the coexistence of two extensive components. In Fig.~\ref{fig:coex_validation_void_L1}, the size of the giant non-occupied component---initially equal to the size of the original giant component $S_\mathrm{cm}$---decreases with increasing $\varphi$ until the component stops being extensive at $\varphi=\bar{\varphi}_c$. Then there is an interval $[\bar{\varphi}_c,\varphi_c]$ where there is no extensive component: the whole graph is fragmented into non-extensive observable islands. The giant occupied component finally emerges at $\varphi=\varphi_c$ and its size increases with increasing $\varphi$ until it is equal to the size of the original giant component. The same behavior is observed in Fig.~\ref{fig:coex_validation_overlap_L1} except that in this case the original giant component is dense enough for the giant occupied component to emerge \textit{before} the giant non-occupied component disappears. Hence whenever $\varphi_c<\bar{\varphi}_c$, there is an interval $[\varphi_c,\bar{\varphi}_c]$ in which two extensive components coexist.

In the context of observability as considered by Yang \textit{et al.}, directly occupied nodes are monitored in such a way that the state of their first neighbors is known as well (case where $L=1$, see the Appendix for a discussion of the case $L=2$) \cite{Yang12_PhysRevLett}. The existence of a giant occupied component then means that a macroscopic contiguous fraction of the graph can be monitored. However, coexistence suggests that monitoring a macroscopic portion of a graph does not prevent a macroscopic event to occur on this graph unbeknown to the observer.
The condition for which there is coexistence is rather simple: the underlying extensive component (the one of the original graph) must be \textit{sufficiently dense} to sustain two giant components. As discussed in Sec.~\ref{sec:coex_real}, this condition is fulfilled in several real systems, with coexistence extending over a wide interval $[\varphi_c,\bar{\varphi}_c]$ in some cases.
%
%
%
%
\section{Mathematical description for arbitrary depth} \label{sec:coex_gen_math_from}
%
The mapping to multitype random graphs can be readily generalized to an arbitrary depth ($L$). The procedure to generate these graphs proceeds initially as for $L=1$, but instead of closing the graph after type 1 nodes have been selected, the remaining free stubs stemming out of type 1 nodes are randomly paired with any free stubs in the whole graph. The nodes thereby reached have either already been tagged as type 1, or have not been tagged and are henceforth considered to be of type 2. The remaining free stubs of type 2 nodes are then randomly paired with any free stubs in the whole graph to determine type 3 nodes. This iterative assignment of node types is repeated until type $L$ nodes are selected. The graph is then finally closed by randomly matching all remaining free stubs; nodes that have not been assigned a type are said to be non-occupied (type $L+1$). In the end, there is a total of $L+2$ node types.

With this iterative assignment of node types in mind, we generalize the mathematical description introduced in the previous section. The probability $\varepsilon_i^{(L)}$ that a random edge leads to a type $i$ node is
\begin{align}
 \varepsilon_i^{(L)} =
  \begin{cases}
   \varphi & i=0 \\
   (1\!-\!\varphi) \big[G_1\big(\chi_{i-1}^{(L)}\big) - G_1\big(\chi_{i}^{(L)}\big) \big] & 1 \leq i \leq L \\
   (1\!-\!\varphi) G_1\big(\chi_{L}^{(L)}\big) & i=L+1
 \end{cases} \label{eq:coex_gen_epsilon} \ ,
\end{align}
where we have defined
\begin{align} \label{eq:coex_gen_chi}
 \chi_i^{(L)} = \begin{cases} 1 & i=0 \\ 1 - \sum_{j=0}^{i-1}\varepsilon_j^{(L)} & i \geq 1 \end{cases} \ .
\end{align}
Similarly, the probability $w_i^{(L)}$ for a random node to be of type $i$ is
\begin{align}
 w_i^{(L)} =
  \begin{cases}
   \varphi & i=0 \\
   (1\!-\!\varphi) \big[G_0\big(\chi_{i-1}^{(L)}\big) - G_0\big(\chi_{i}^{(L)}\big) \big] & 1 \leq i \leq L \\
   (1\!-\!\varphi) G_0\big(\chi_{L}^{(L)}\big) & i=L+1
 \end{cases} \label{eq:coex_gen_w} \ .
\end{align}
The value of $\varepsilon_0^{(L)}$ and of $w_0^{(L)}$ come from the definition of depth-$L$ percolation itself, that is that type 0 nodes are assigned randomly with probability $\varphi$. Using Eqs.~\eqref{eq:coex_gen_epsilon}, we see that $\chi_{i}^{(L)} = (1-\varphi)G_1\big(\chi_{i-1}^{(L)}\big)$ meaning that $\chi_i^{(L)}$ is the probability that the type of the node at the end of a random edge is not lower than $i$. Hence for $1\leq i\leq L$, the values of $\varepsilon_i^{(L)}$ and of $w_i^{(L)}$ equal to the probability that the type of the node is not lower than $i-1$ minus the probability that its type is not lower than $i$. The value of $\varepsilon_{L+1}^{(L)}$ and of $w_{L+1}^{(L)}$ follow directly. Both sets of probabilities are normalized.

We compute the joint degree distribution of each node type in a similar manner. Based on the procedure described above, type $i$ nodes are randomly and independently connected: (i) to no node whose type is lower than $i-1$, (ii) to at least one type $(i-1)$ nodes with probability $\varepsilon_{i-1}^{(L)}$, (iii) to type $i$ nodes with probability $\varepsilon_{i}^{(L)}$, (iv) to type $(i+1)$ nodes with the complementary probability $\chi_{i+1}^{(L)}$, and (v) and to no node whose type is higher than $i+1$. %
\begin{figure*}[t]
  \subfigure[\ power-law with $\alpha = 2.5$ and $\kappa = 75$]{\label{fig:coex_variousL_1} \includegraphics[width=0.45\textwidth]{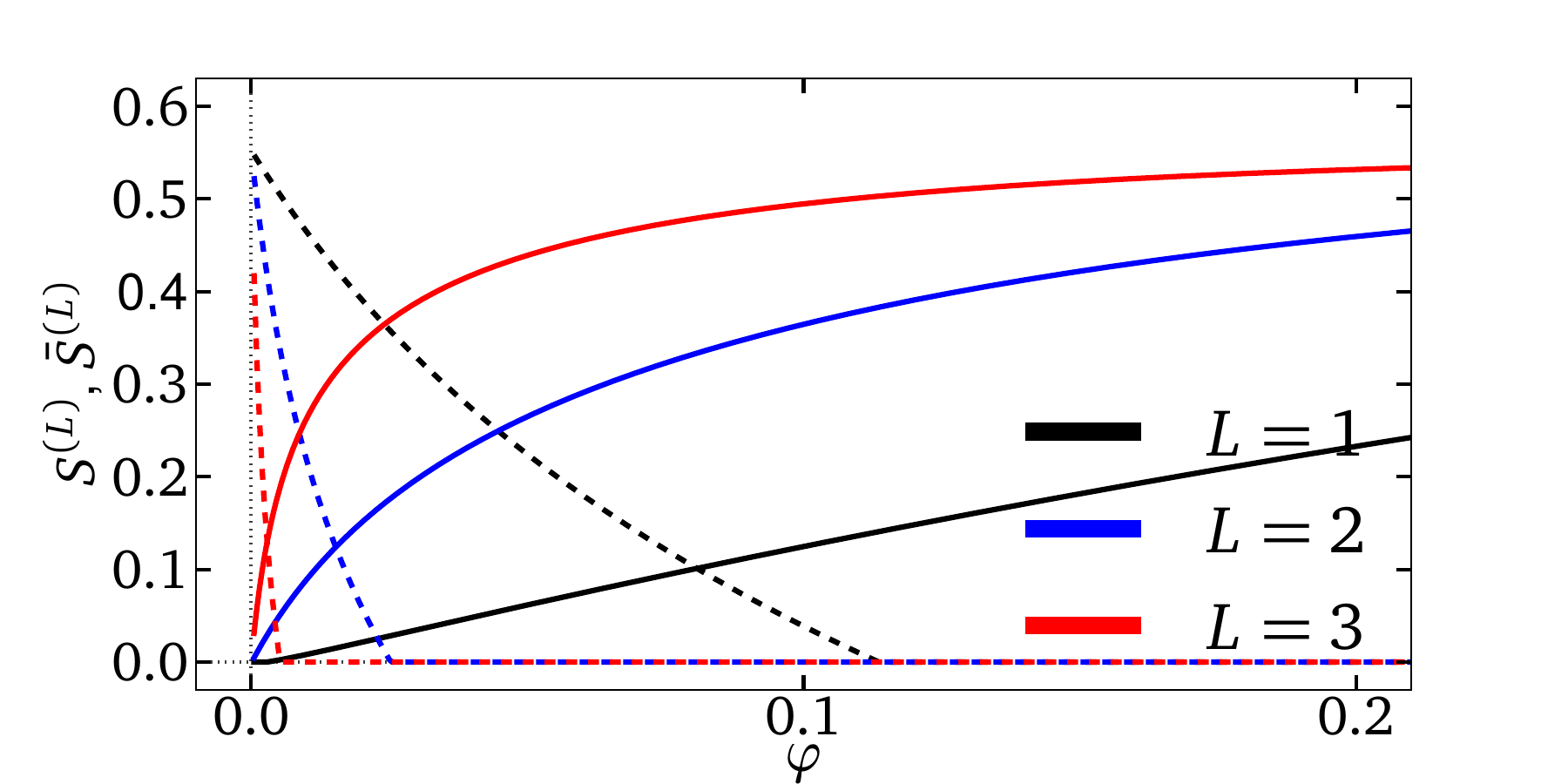}}
  \subfigure[\ exponential with $\langle k \rangle = 2.8$]{\label{fig:coex_variousL_2} \includegraphics[width=0.45\textwidth]{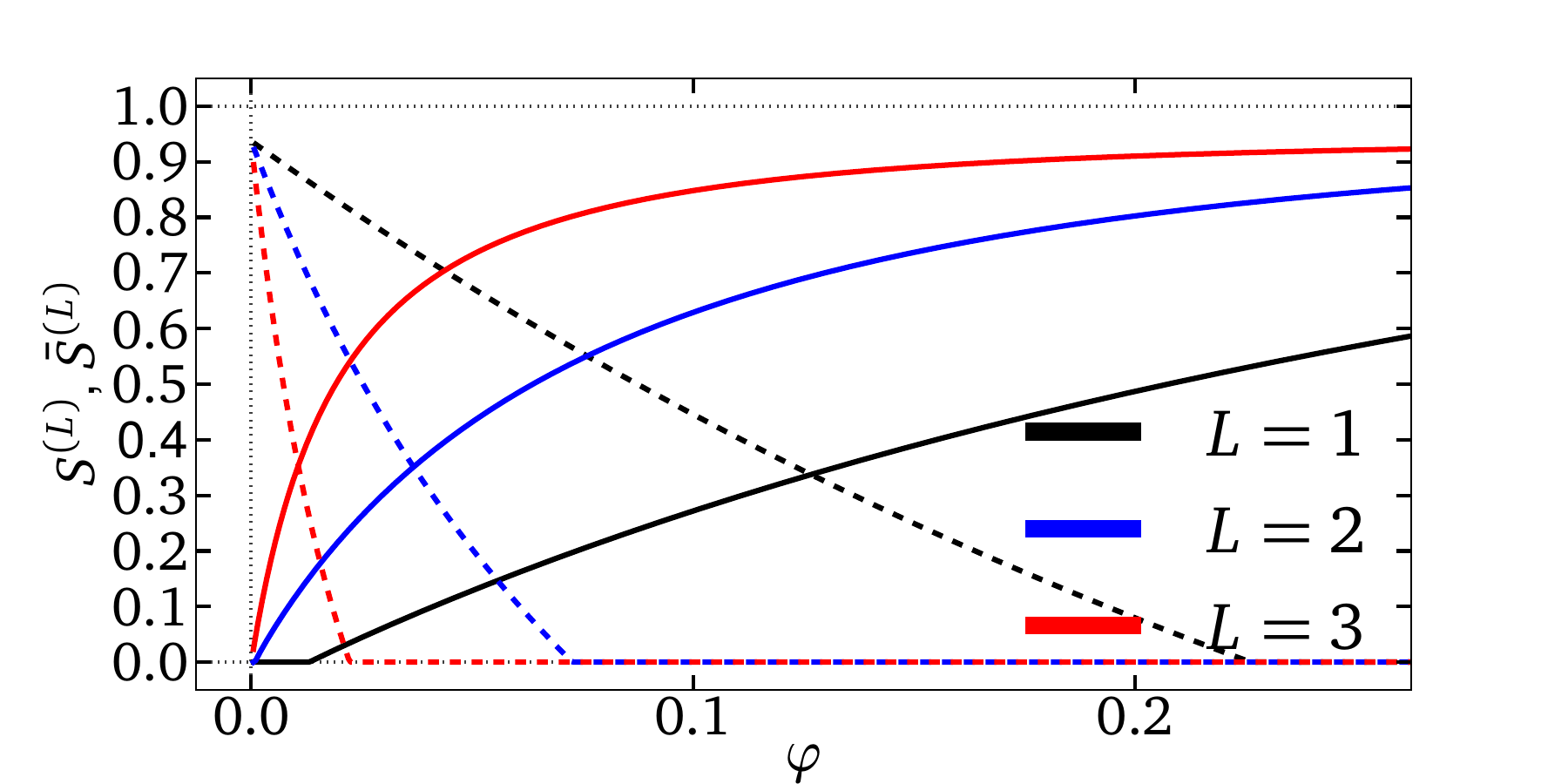}} \\
  \subfigure[\ power-law with $\alpha = 2.5$ and $\kappa = 7.5$]{\label{fig:coex_variousL_3} \includegraphics[width=0.45\textwidth]{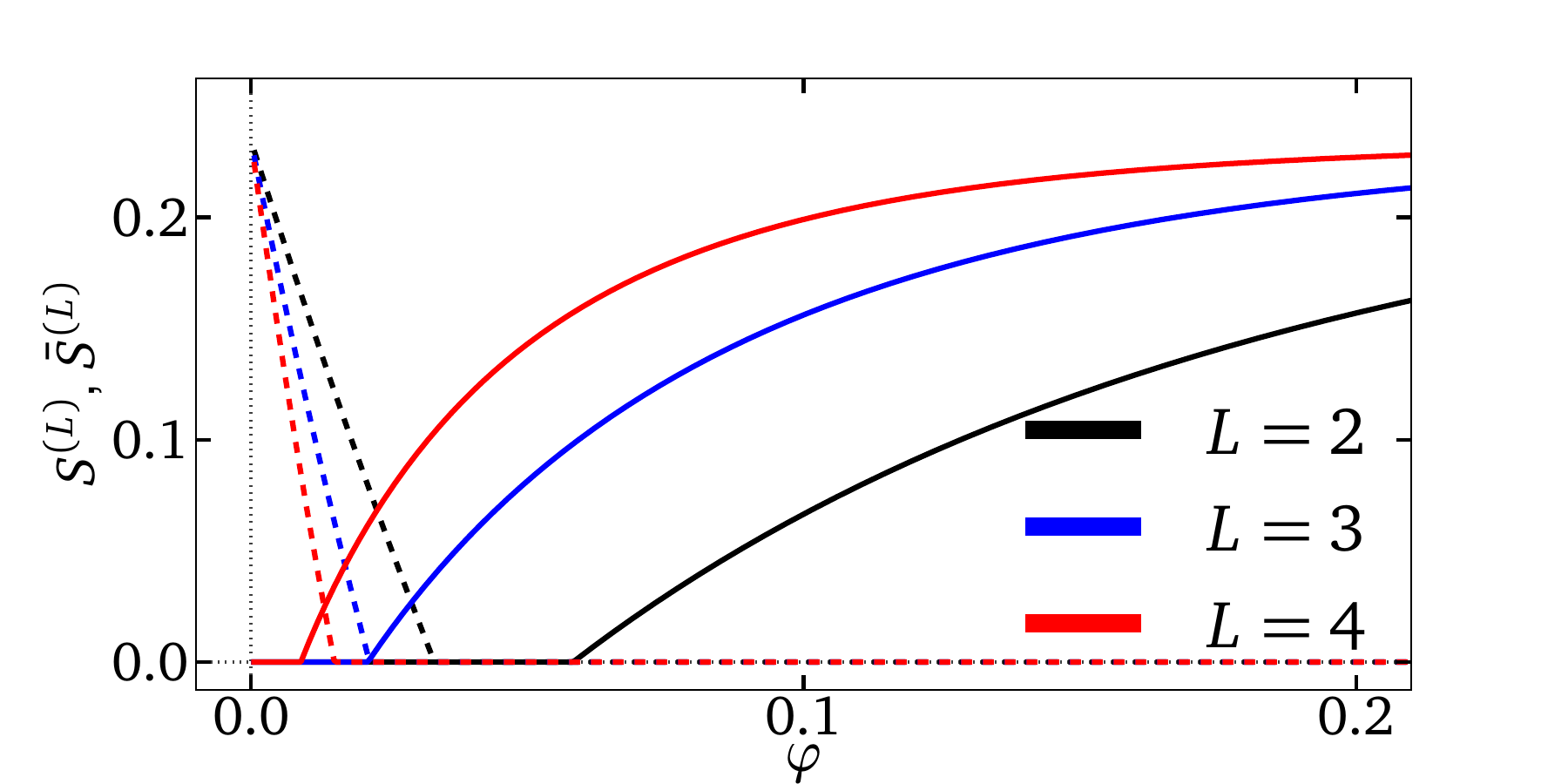}}
  \subfigure[\ exponential with $\langle k \rangle = 1.6$]{\label{fig:coex_variousL_4} \includegraphics[width=0.45\textwidth]{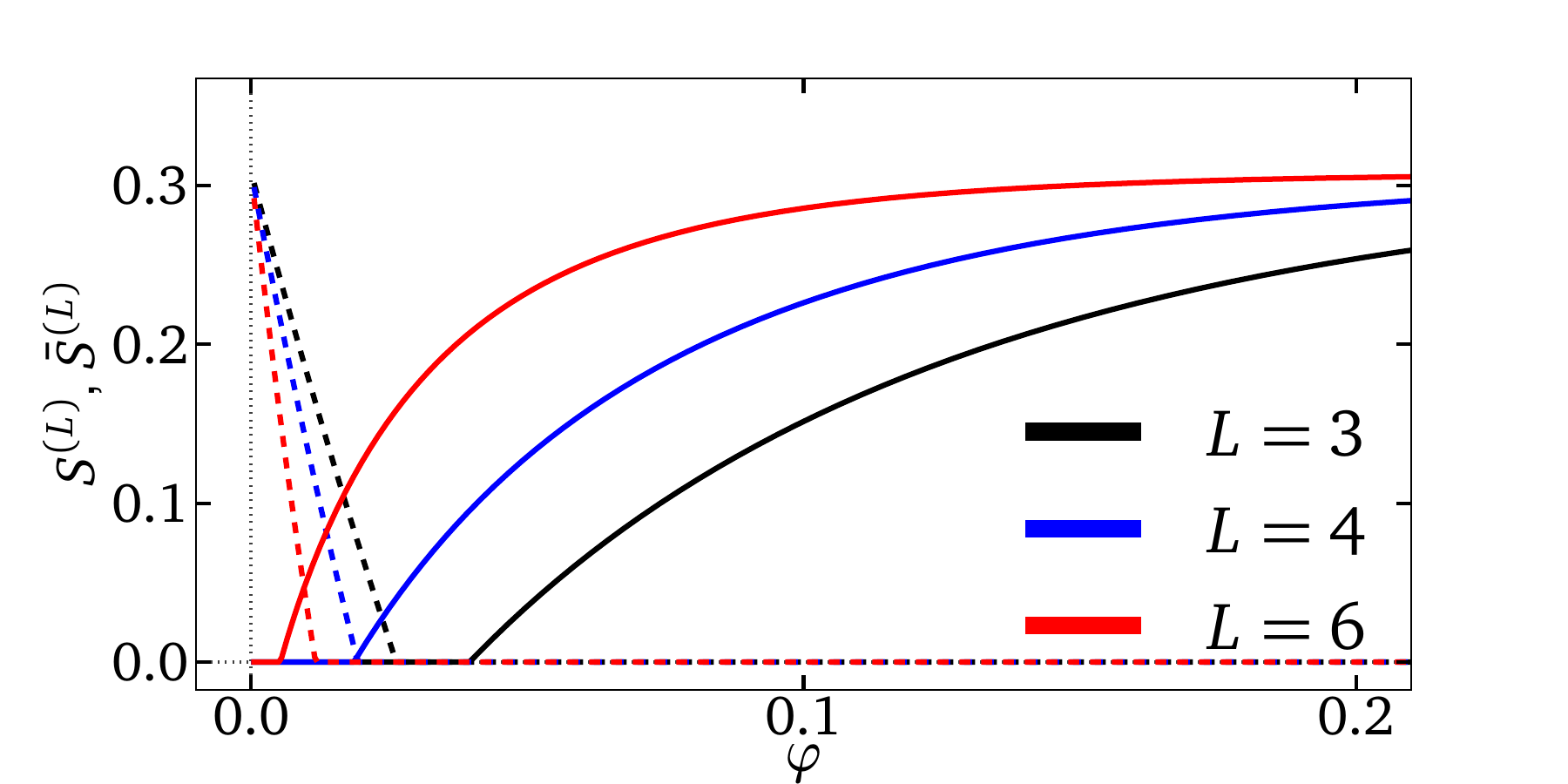}} \\
  \subfigure[\ Poisson with $\langle k \rangle = 1.2$]{\label{fig:coex_variousL_5} \includegraphics[width=0.45\textwidth]{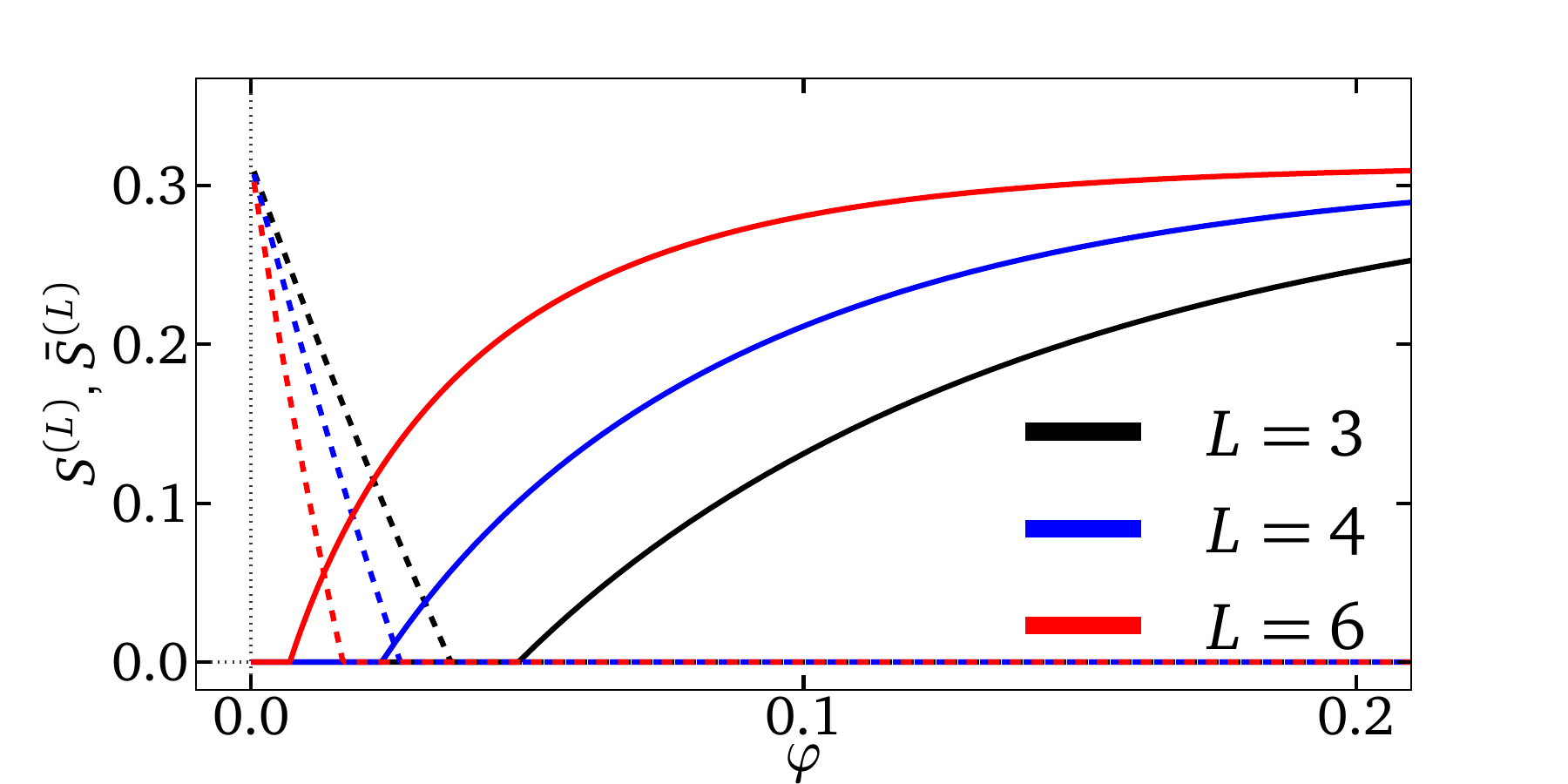}}
  \subfigure[\ Thresholds vs. depth]{\label{fig:coex_threshold_vs_depth} \includegraphics[width=0.45\textwidth]{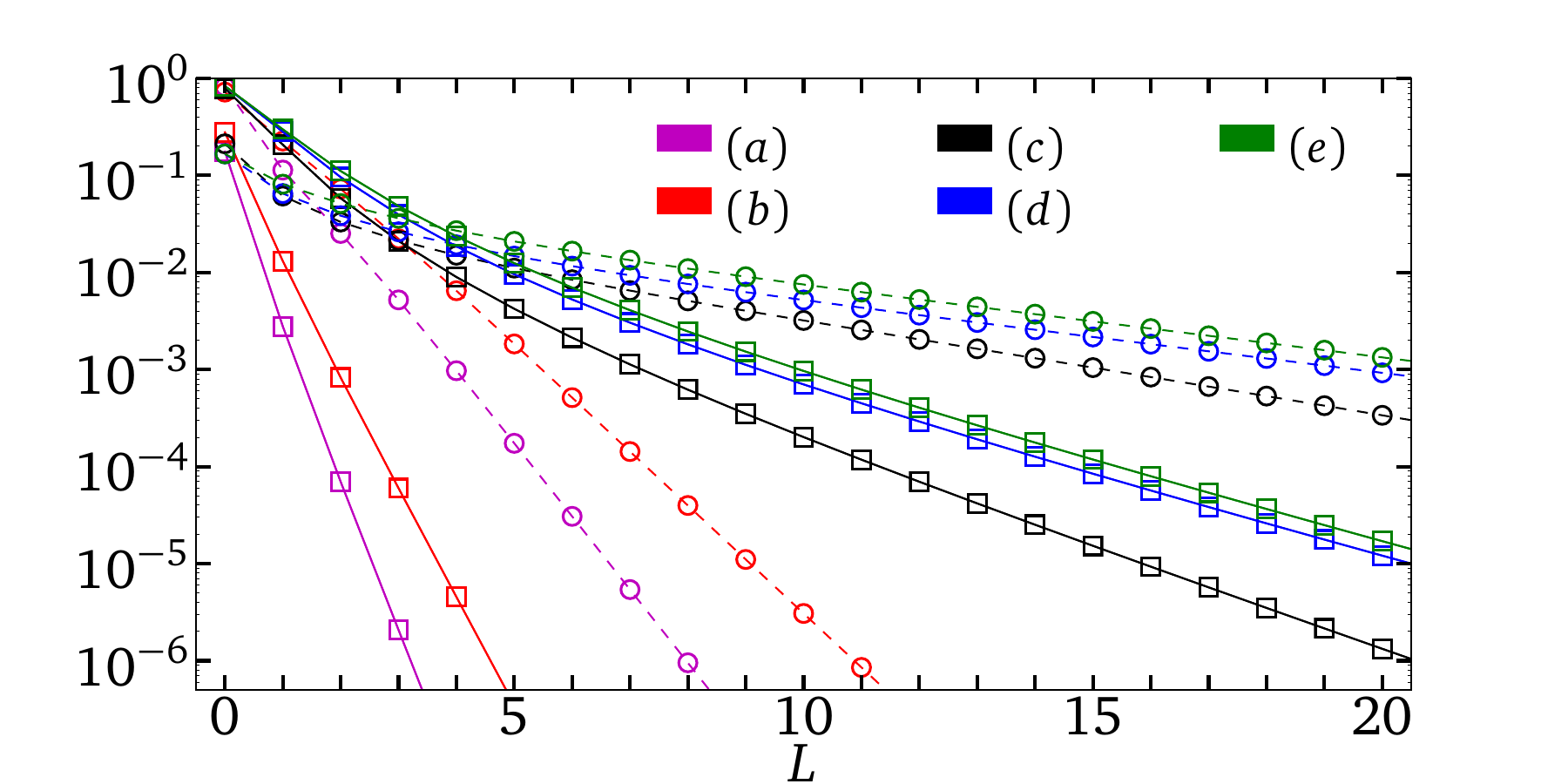}}
  \caption{\label{fig:coex_variousL}(color online). Effect of varying the depth $L$ on the coexistence regime. \subref{fig:coex_variousL_1}--\subref{fig:coex_variousL_5} The size of the non-occupied giant component ($\bar{S}^{(L)}$, dash lines) and of the occupied giant component ($S^{(L)}$, solid lines) are shown as a function of $\varphi$ for different values of the depth $L$ and different degree distributions. The power-law degree distribution is defined as $P(k) = k^{-\alpha}\mathrm{e}^{-k/\kappa}/\mathrm{Li_{\alpha}(\mathrm{e}^{1/\kappa})}$ with $k \geq 1$ and $\mathrm{Li}_{\alpha}(x)$ denoting the polylogarithm. The Poisson degree distribution is defined as $P(k) = \lambda^k \mathrm{e}^{-\lambda}/k!$ with $k\geq0$ and $\lambda=\langle k \rangle$. See the caption of Fig.~\ref{fig:coex_validation} for the definition of the exponential degree distribution. All curves were obtained by solving Eqs.~\eqref{eq:coex_gen_goc_S}--\eqref{eq:coex_gen_gnc_a}. Figures~\subref{fig:coex_variousL_1}--\subref{fig:coex_variousL_5} are a representative subset of the behaviors obtained with many realistic and commonly used degree distributions. \subref{fig:coex_threshold_vs_depth} Behavior of $\bar{\varphi}_c^{(L)}$ (circles) and $\varphi_c^{(L)}$ (squares) as a function of $L$ using the degree distributions of \subref{fig:coex_variousL_1}--\subref{fig:coex_variousL_5}. Values were obtained from \eqref{eq:coex_gen_gnc_threshold}, and by analyzing the stability of Eqs.~\eqref{eq:coex_gen_goc_a} around $\bm{a}=\bm{1}$. Lines have been added to guide the eye.} 
\end{figure*}
Enforcing the normalization of the resulting joint degree distributions, we obtain the following associated pgf
\begin{widetext}
\begin{align}
 g_i^{(L)}(\bm{x}) & = 
   \begin{cases}
     G_0\Big(\varphi x_{0,0} + [1-\varphi]x_{0,1}\Big) & i=0 \\
     \displaystyle\frac{G_0\Big( \varepsilon_{i-1}^{(L)}x_{i,i-1} + \varepsilon_{i}^{(L)}x_{i,i} + \chi_{i+1}^{(L)}x_{i,i+1}\Big) - G_0\Big( \varepsilon_{i}^{(L)}x_{i,i} + \chi_{i+1}^{(L)}x_{i,i+1}\Big)}{G_0\Big( \chi_{i-1}^{(L)}\Big) - G_0\Big( \chi_{i}^{(L)}\Big)} & 1 \leq i \leq L \\
     \displaystyle \frac{G_0\Big( \varepsilon_{L}^{(L)}x_{L+1,L} + \varepsilon_{L+1}^{(L)}x_{L+1,L+1}\Big)}{G_0\Big( \chi_{L}^{(L)}\Big)} & i=L+1 \\
   \end{cases} \label{eq:coex_gen_g} \ .
\end{align}
\end{widetext} 

Following Ref.~\cite{Allard09_PhysRevE}, we set $x_{L,L+1}=x_{L+1,L}=1$ in Eqs.~\eqref{eq:coex_gen_g} and the relative size of the giant occupied component, $S^{(L)}$, is computed from
\begin{align} \label{eq:coex_gen_goc_S}
  S^{(L)} & = \sum_{i=0}^{L} w_i^{(L)} \big[1 - g_i^{(L)}(\bm{a}^{(L)}) \big] 
\end{align}
where $\bm{a}^{(L)}\equiv\{a_{ij}^{(L)}\}_{i,j=0,\ldots,L}$ is the fixed point---the smallest solution in $[0,1]^{(L+1)^2}$---of the system of equations
\begin{align} \label{eq:coex_gen_goc_a}
  a_{ij}^{(L)} & = \frac{\partial g_{j}^{(L)}(\bm{a}^{(L)})/\partial x_{ji}}{\partial g_{j}^{(L)}(\bm{1})/\partial x_{ji}}
\end{align}
\begin{figure*}[t]
  \subfigure[\ arXiv co-authorship network]{\label{fig:coex_real_arXiv} \includegraphics[width=0.45\textwidth]{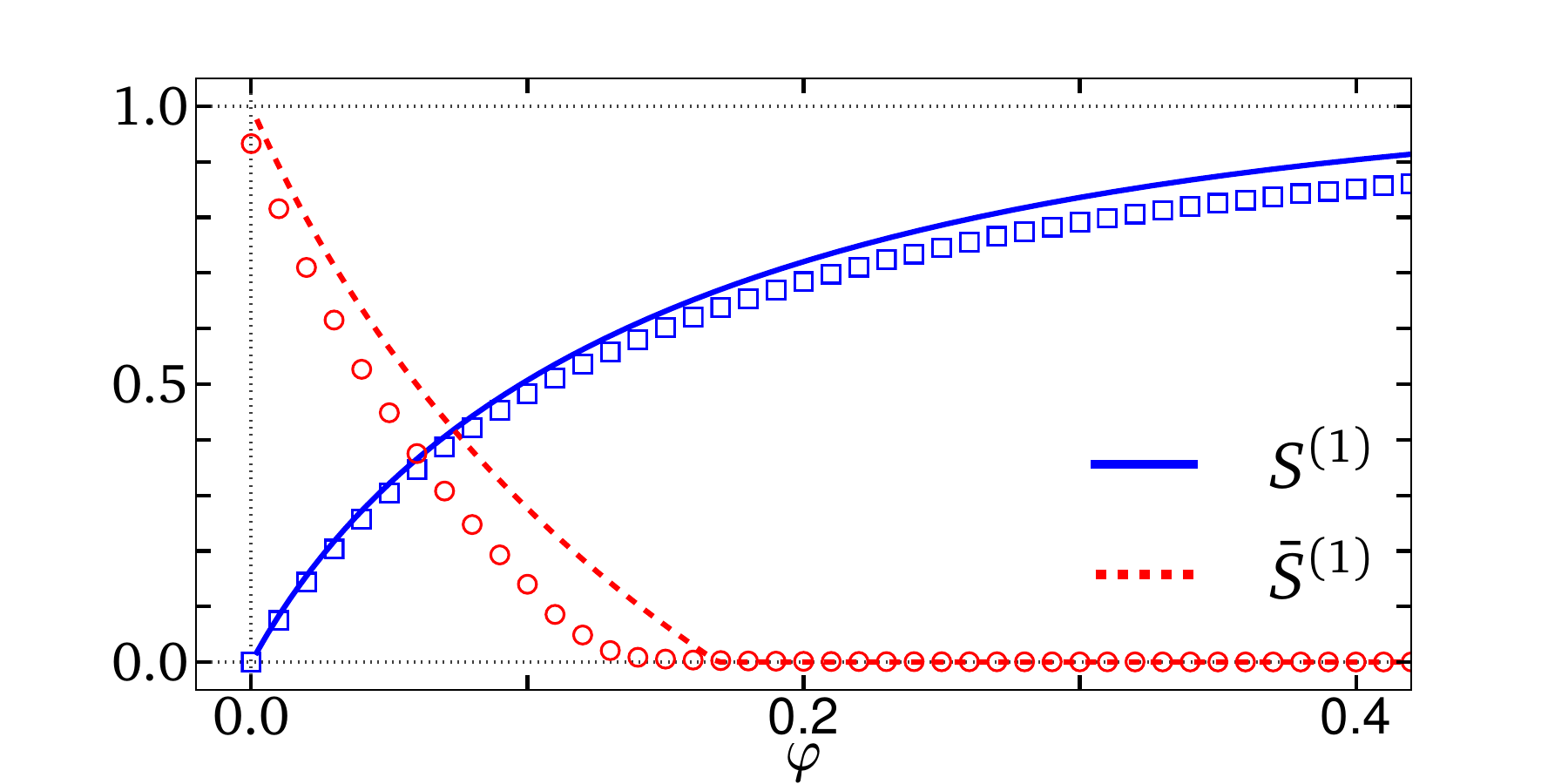}}
  \subfigure[\ Email communication network]{\label{fig:coex_real_Email} \includegraphics[width=0.45\textwidth]{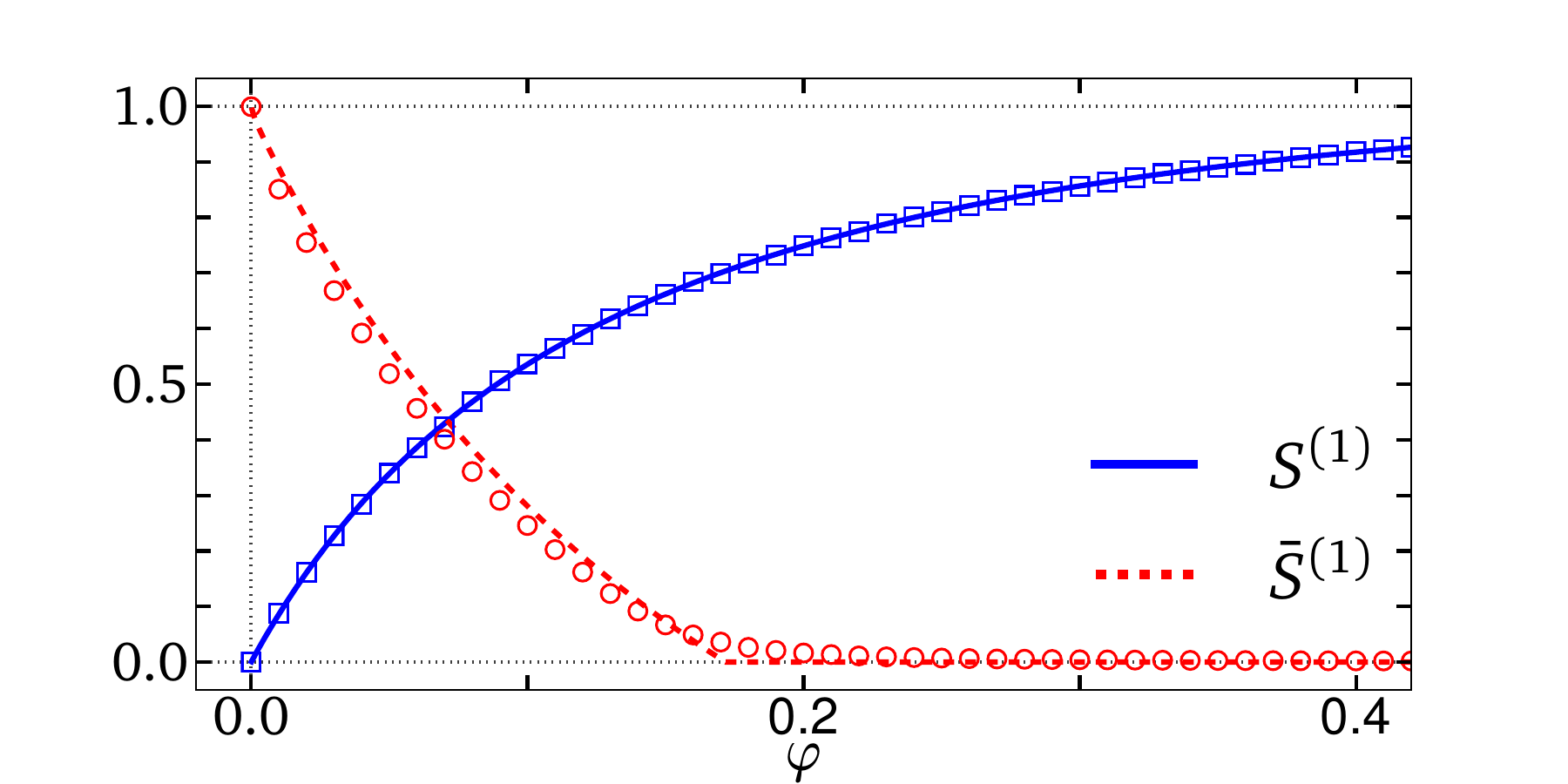}} \\
  \subfigure[\ Internet]{\label{fig:coex_real_Internet} \includegraphics[width=0.45\textwidth]{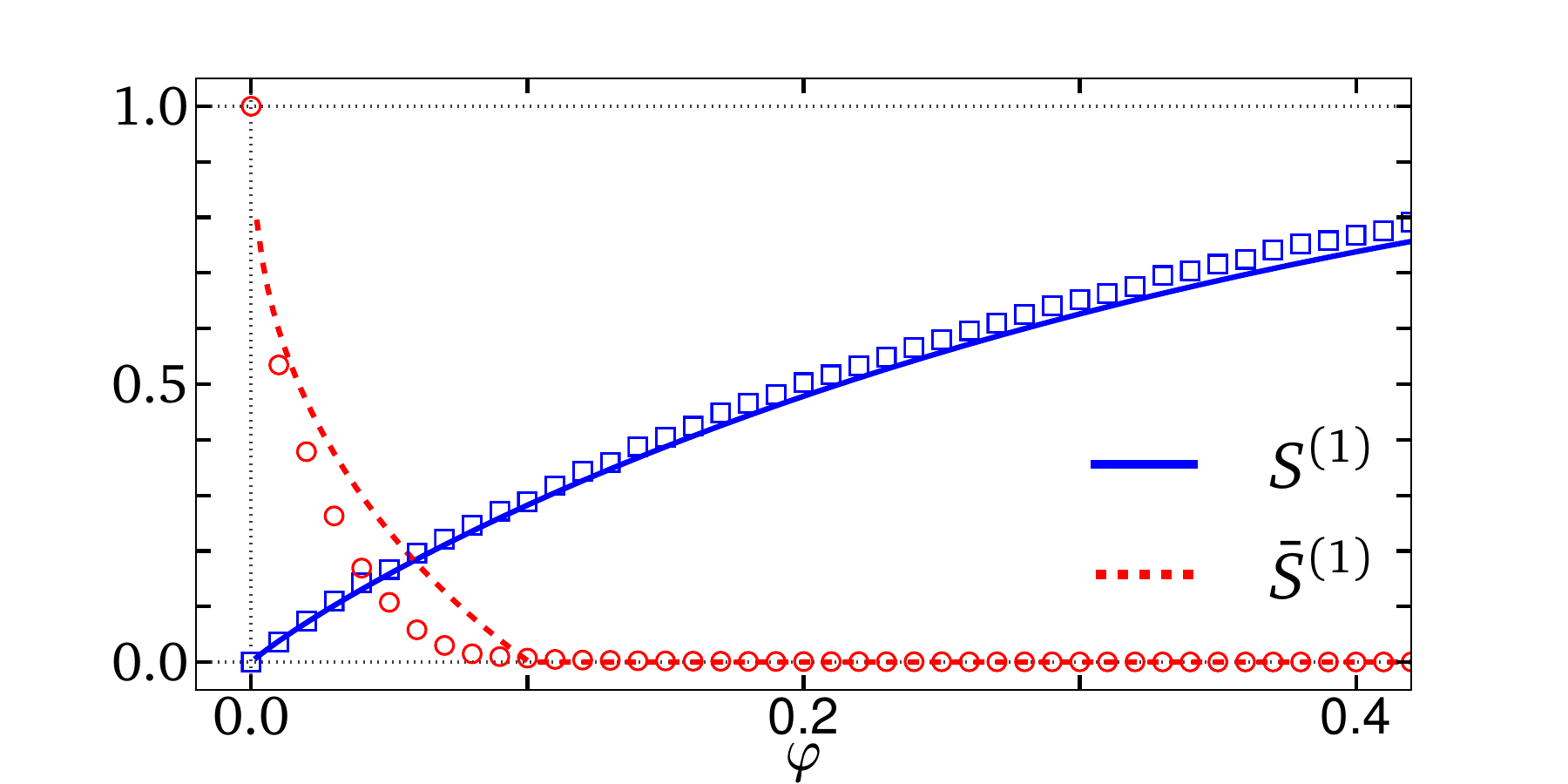}}
  \subfigure[\ Protein interaction network]{\label{fig:coex_real_ProteinCore} \includegraphics[width=0.45\textwidth]{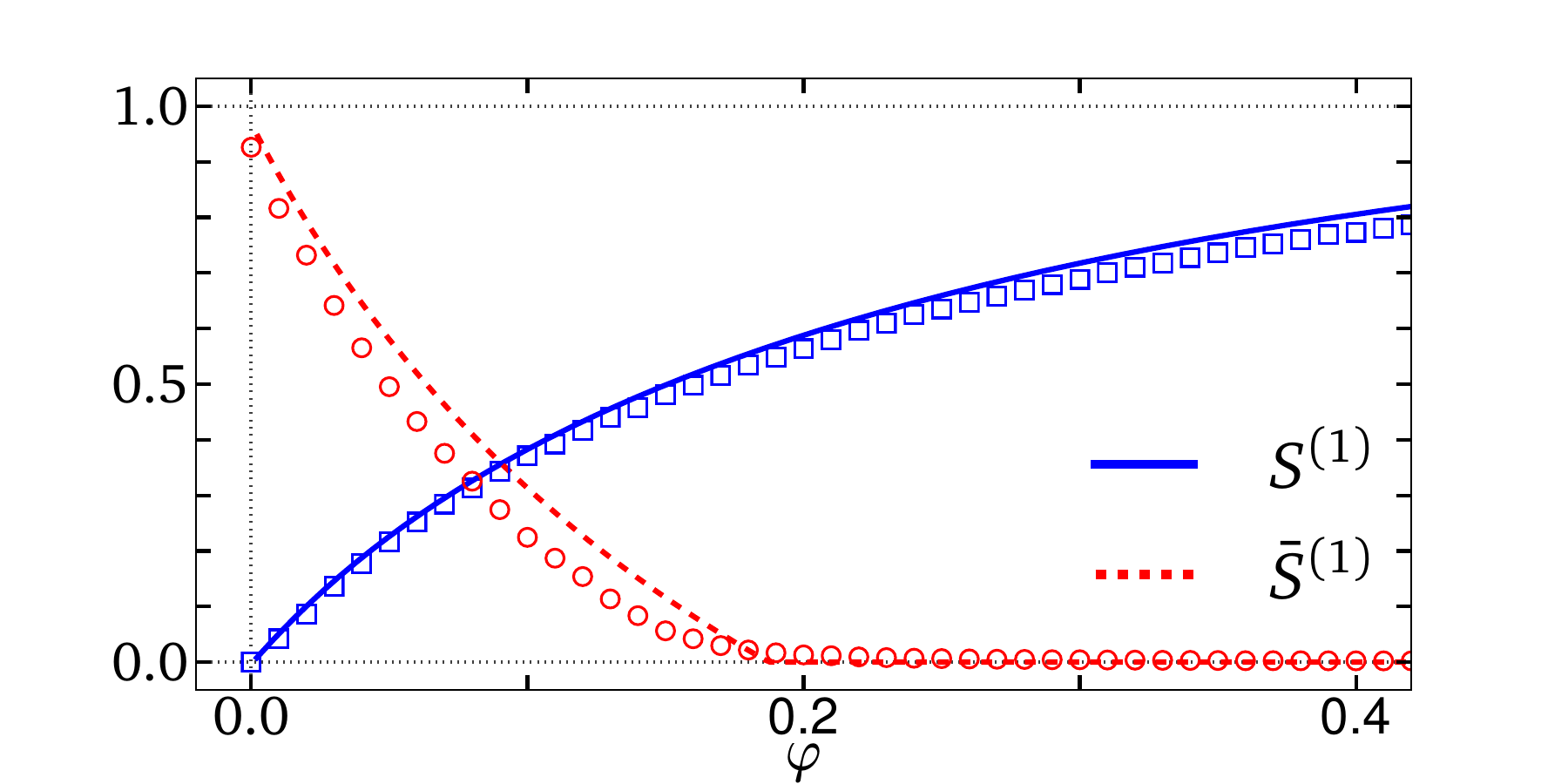}} \\
  \subfigure[\ Slashdot online social network]{\label{fig:coex_real_Slashdot} \includegraphics[width=0.45\textwidth]{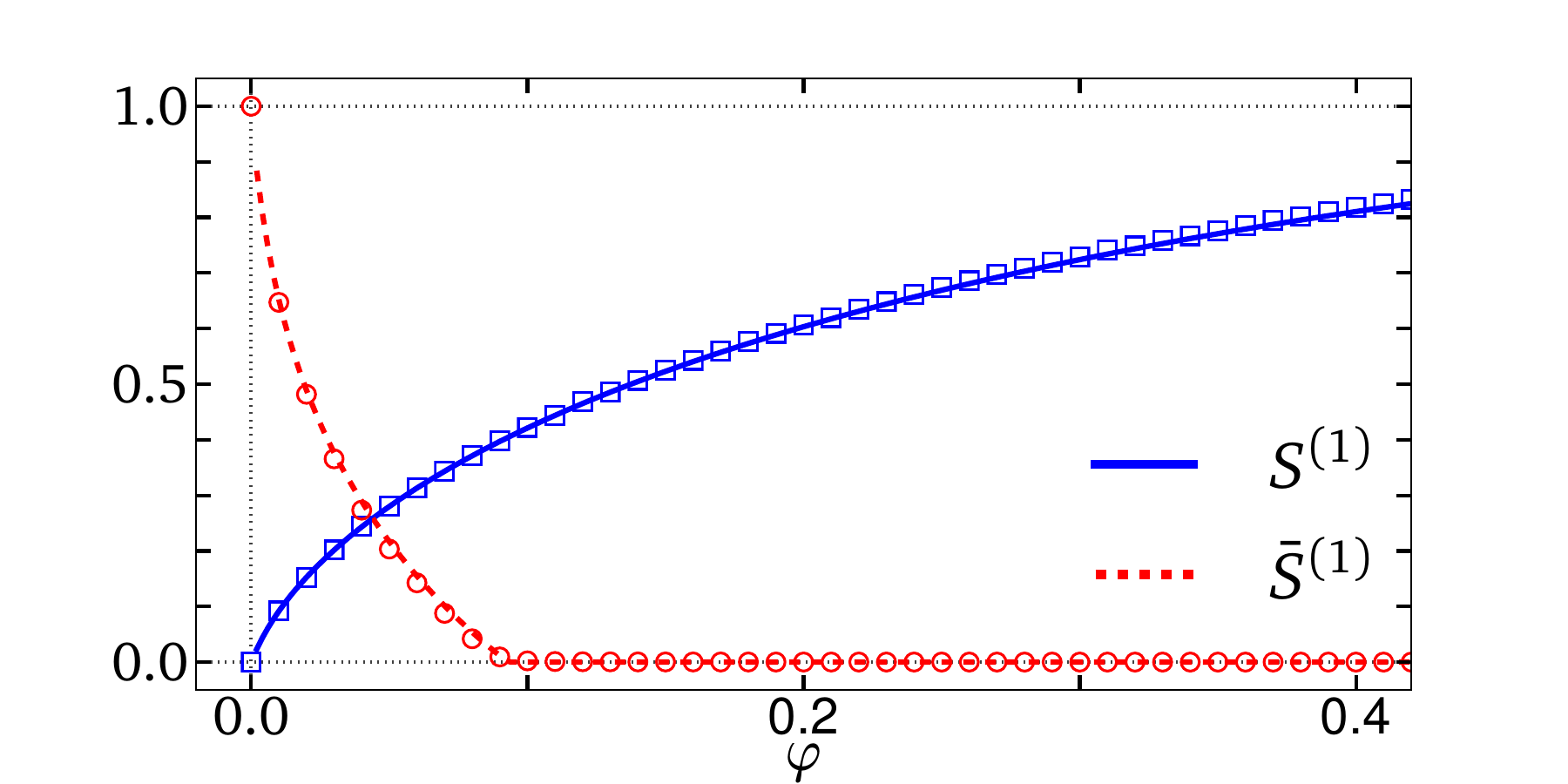}}
  \subfigure[\ Degree distributions]{\label{fig:coex_real_dd_powlaw} \includegraphics[width=0.45\textwidth]{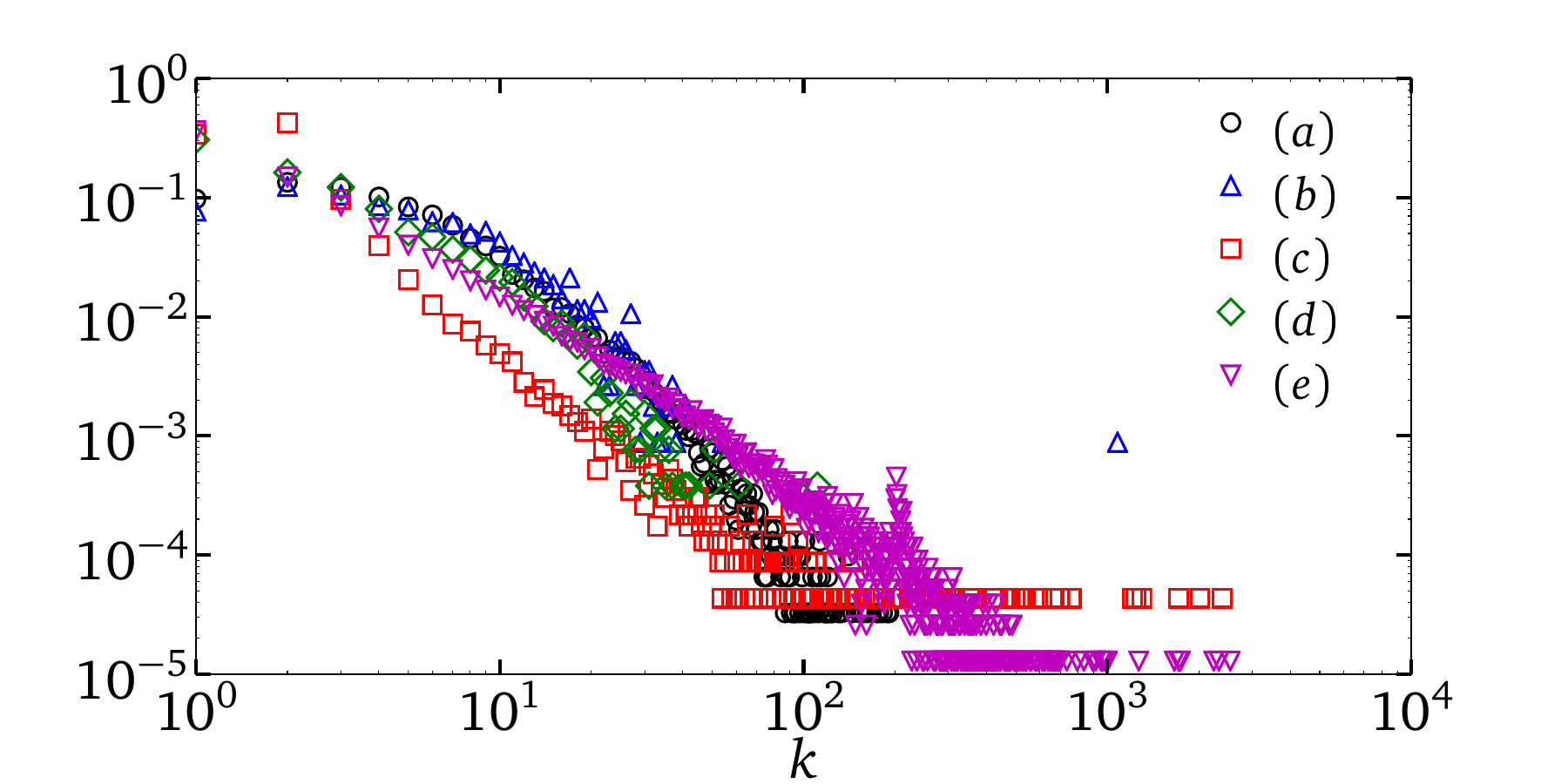}} \\
  \caption{\label{fig:coex_real_nongeo}(color online). \subref{fig:coex_real_arXiv}--\subref{fig:coex_real_Slashdot} Depth-1 percolation on graphs extracted from real non-geographically-constrained complex systems (see Table~\ref{tab:coex_databases}). Symbols represent the average (100 simulations minimum) relative size of the largest occupied ($S^{(1)}$) and non-occupied ($\bar{S}^{(1)}$) components found in these graphs where directly occupied nodes were selected randomly with probability $\varphi$. Lines were obtained by solving Eqs.~\eqref{eq:coex_goc_a}, \eqref{eq:coex_goc_S}, \eqref{eq:coex_gnc_a} and \eqref{eq:coex_gnc_S} with the degree distribution extracted from each graph [shown in \subref{fig:coex_real_dd_powlaw}].}
\end{figure*}%
with $i,j=0,\ldots,L$. The point at which the giant occupied component emerges, $\varphi_c^{(L)}$, is obtained by a linear stability analysis of the fixed point $\{a_{ij}\}=\bm{1}$ with $i,j=0,\ldots,L$. Although the corresponding Jacobian matrix is composed of recurrent patterns of non-zero elements---due to the hierarchy of node types---it has not been possible to extract a useful general equation for $\varphi_c^{(L)}$. The relative size of the non-occupied component, $\bar{S}^{(L)}$, is computed from
\begin{align} \label{eq:coex_gen_gnc_S}
  \bar{S}^{(L)} & = w_{L+1} \left[ 1 - \frac{G_0\Big( \varepsilon_{L}^{(L)} + \varepsilon_{L+1}^{(L)}a_{L+1,L+1}^{(L)}\Big)}{G_0\Big( \chi_{L}^{(L)}\Big)} \right] \ ,
\end{align}
where $a_{L+1,L+1}^{(L)}$ is the fixed point of
\begin{align} \label{eq:coex_gen_gnc_a}
  a_{L+1,L+1}^{(L)} & = \frac{G_1\Big( \varepsilon_{L}^{(L)} + \varepsilon_{L+1}^{(L)}a_{L+1,L+1}^{(L)}\Big)}{G_1\Big( \chi_{L}^{(L)}\Big)} \ .
\end{align}
Analyzing the stability of the fixed point $a_{L+1,L+1}=1$, we find that the related critical point, $\bar{\varphi}_c^{(L)}$, is the solution of
\begin{align} \label{eq:coex_gen_gnc_threshold}
 (1-\bar{\varphi}_c^{(L)}) G_1^\prime\big(\chi_L^{(L)}) = 1 \ .
\end{align}
Predictions of Eqs.~\eqref{eq:coex_gen_goc_S}--\eqref{eq:coex_gen_gnc_threshold} are validated in Fig.~\ref{fig:coex_validation_overlap_L3}. Equations derived in Sec.~\ref{sec:coex_depthL_perco} are retrieved directly by setting $L=1$ in Eqs.~\eqref{eq:coex_gen_epsilon}--\eqref{eq:coex_gen_gnc_threshold}. A very accurate approximation of Eqs.~\eqref{eq:coex_gen_gnc_S} for the case $L=2$ has been given in the Supplemental Material provided with Ref.~\cite{Yang12_PhysRevLett}. This case is much more delicate than the case $L=1$: a complete Appendix is devoted to working out the correspondence of the approach of Ref.~\cite{Yang12_PhysRevLett} with the exact calculation provided in this section.
\begin{figure*}[t]
  \subfigure[\ Road network of Pennsylvania]{\label{fig:coex_real_PennRoad} \includegraphics[width=0.45\textwidth]{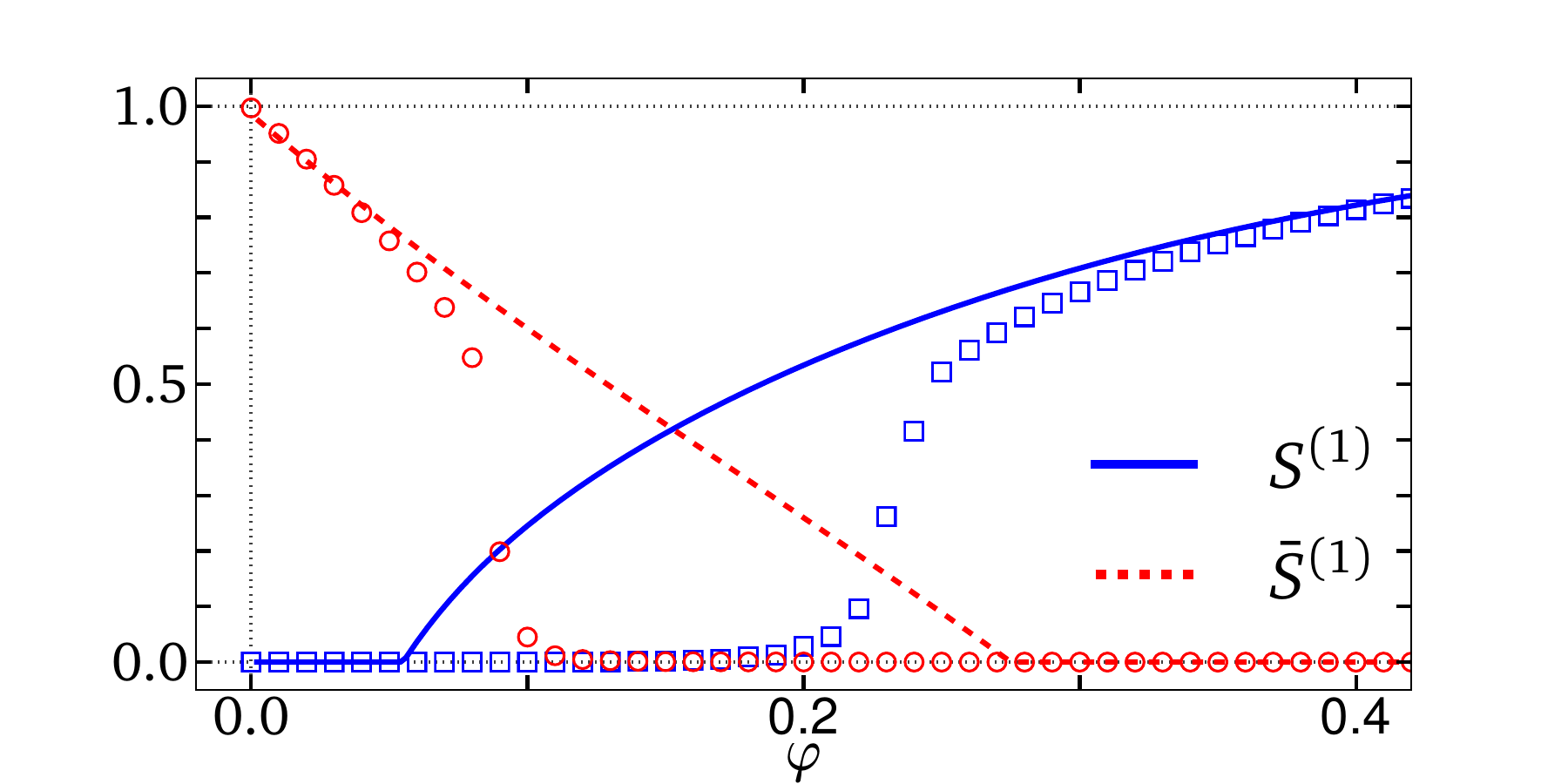}}
  \subfigure[\ Polish power grid]{\label{fig:coex_real_PolishGrid} \includegraphics[width=0.45\textwidth]{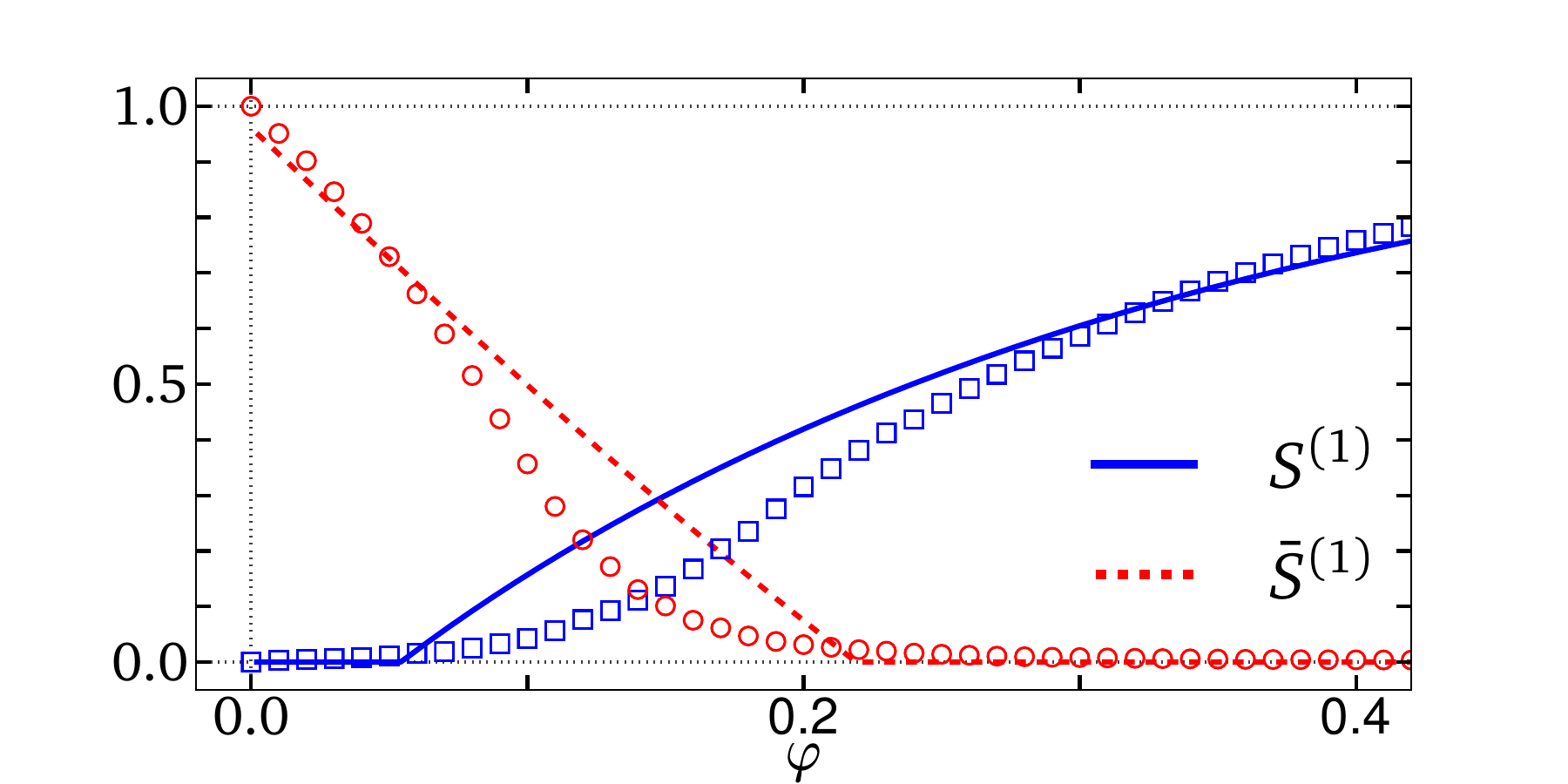}} \\
  \subfigure[\ Degree distributions]{\label{fig:coex_real_dd_expo} \includegraphics[width=0.45\textwidth]{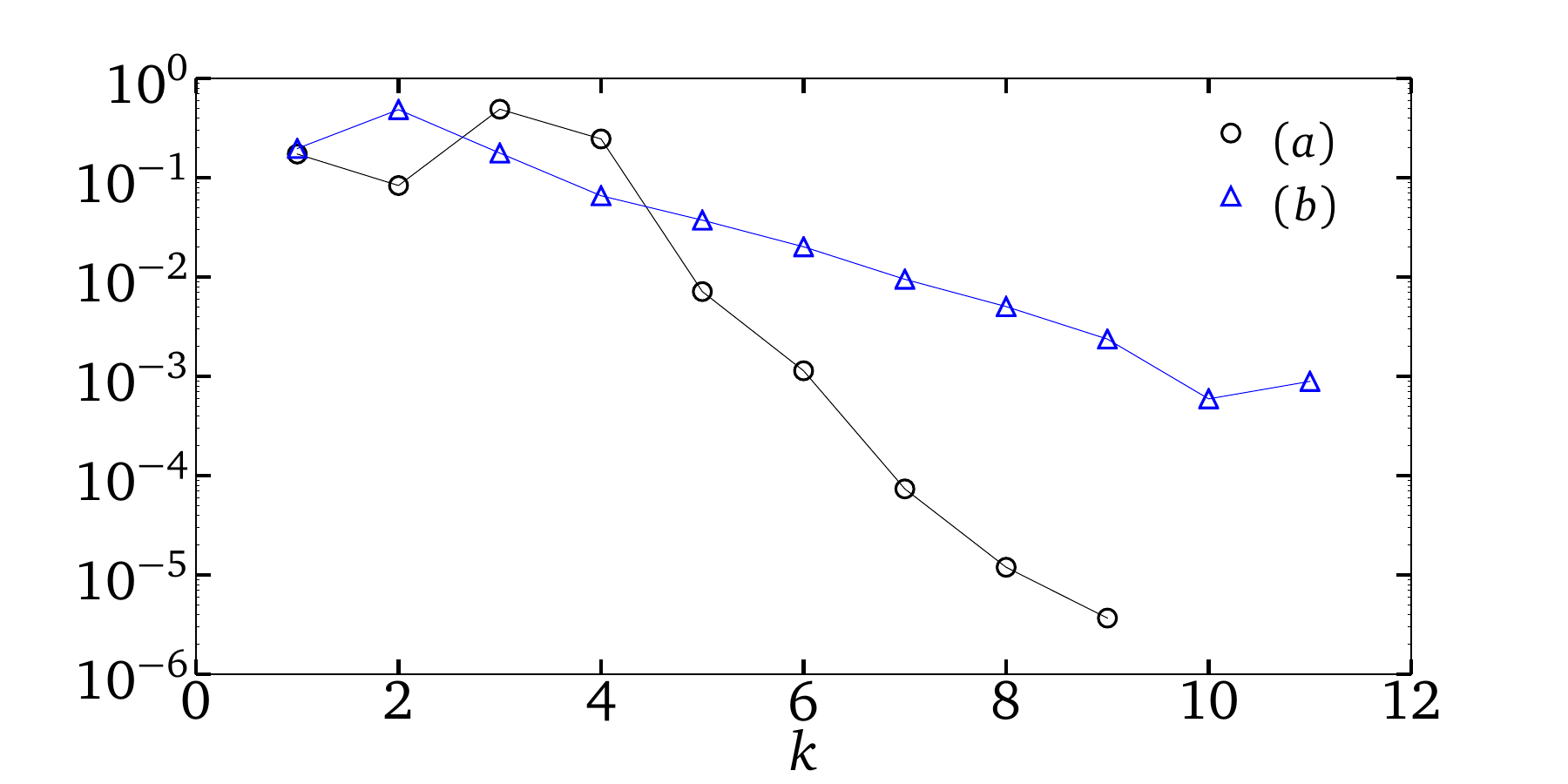}}
  \subfigure[\ Lattices]{\label{fig:coex_real_lattice} \includegraphics[width=0.45\textwidth]{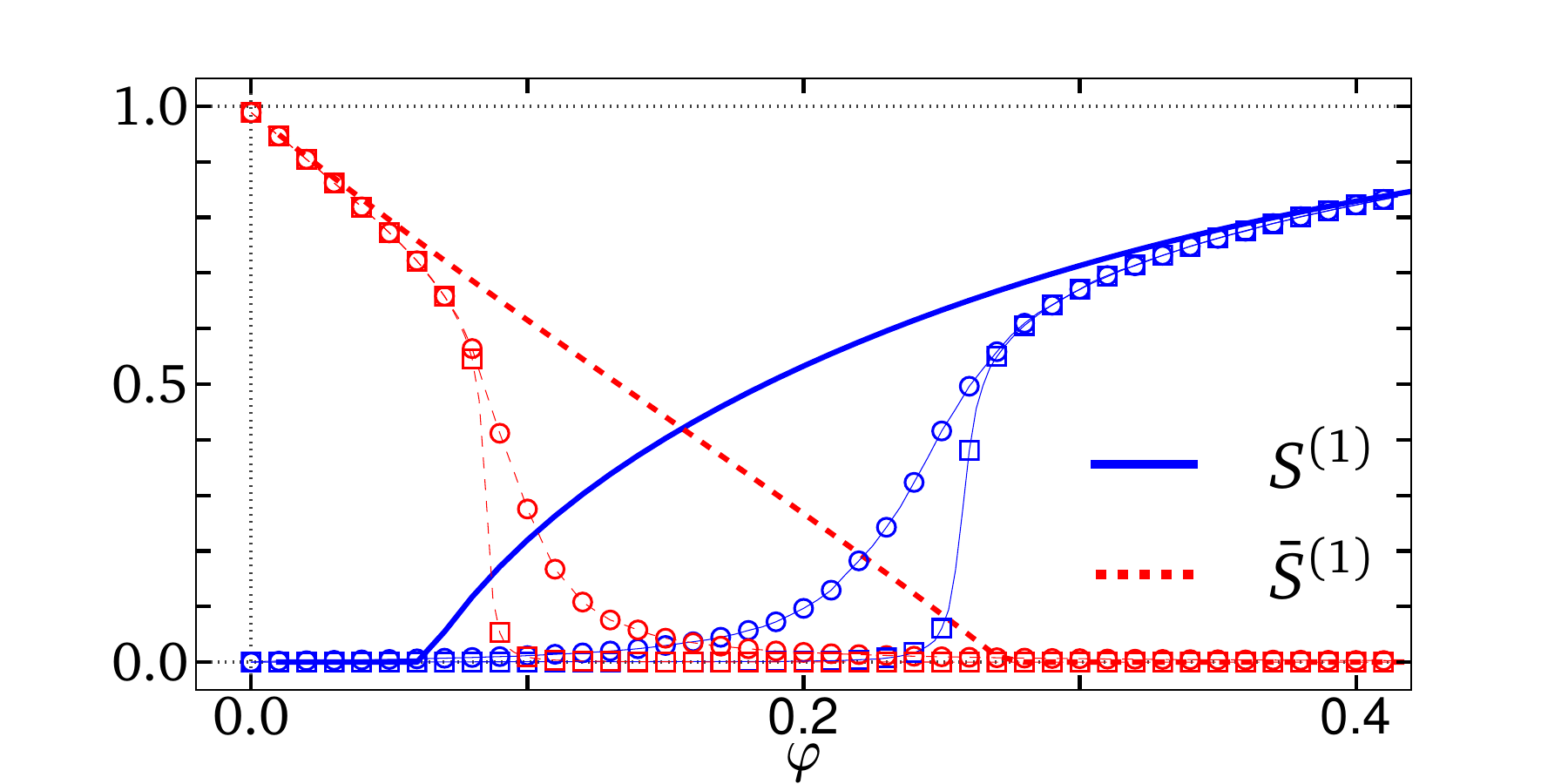}} \\
  \caption{\label{fig:coex_real_geo}(color online). \subref{fig:coex_real_PennRoad}--\subref{fig:coex_real_PolishGrid}Depth-1 percolation on graphs extracted from real geographically-constrained complex systems (see Table~\ref{tab:coex_databases}). Symbols represent the average (100 simulations minimum) relative size of the largest occupied ($S^{(1)}$) and non-occupied ($\bar{S}^{(1)}$) components found in these graphs where directly occupied nodes have been selected randomly with probability $\varphi$. Lines were obtained by solving Eqs.~\eqref{eq:coex_goc_a}, \eqref{eq:coex_goc_S}, \eqref{eq:coex_gnc_a} and \eqref{eq:coex_gnc_S} with the degree distribution extracted from each graph [shown in \subref{fig:coex_real_dd_expo}]. \subref{fig:coex_real_lattice} Depth-1 percolation on square $L \times L$ lattices (circles: $L=70$, squares: $L=1000$) where edges are randomly removed with probability $p=0.30$ ($\langle k \rangle = 2.8$). Symbols represent the average (100 simulations minimum) relative size of the largest occupied ($S^{(1)}$) and non-occupied ($\bar{S}^{(1)}$) components. Lines (with no symbols) are the predictions of Eqs.~\eqref{eq:coex_goc_a}, \eqref{eq:coex_goc_S}, \eqref{eq:coex_gnc_a} and \eqref{eq:coex_gnc_S} with the binomial degree distribution $P(k) = \binom{4}{k}(1-p)^k p^{4-k}$ with $0 \leq k \leq 4$. Lines have been added between symbols in \subref{fig:coex_real_dd_expo}--\subref{fig:coex_real_lattice} to guide the eye.}
\end{figure*}
%
%
%
\subsection{The symmetric case $L=0$} \label{sec:coex_case_L0}
%
The case $L=0$ corresponds to traditional site percolation on random graphs. In the context of observability, it is somewhat trivial as it is symmetric: the non-occupied giant component behaves exactly as the occupied one under the substitution $\varphi \rightarrow 1-\varphi$. It is however an interesting case as expressions for $\varphi_c^{(0)}$ and $\bar{\varphi}_c^{(0)}$ can be obtained in closed form
\begin{align} \label{eq:coex_si_thresholds_L0}
 \varphi_c^{(0)} & = 1 - \bar{\varphi}_c^{(0)} = \frac{1}{G_1^\prime(1)} \ .
\end{align}
As expected, this corresponds to the threshold value obtained for site percolation on random graphs \cite{Callaway00_PhysRevLett}. Asking for the coexistence of the two extensive components (i.e., $\varphi_c^{(0)}<\bar{\varphi}_c^{(0)}$), we find the condition
\begin{align*}
  G_1^\prime(1) > 2 \ .
\end{align*}
This offers a quantitative criterion for the original giant component to be \textit{dense enough} to sustain coexistence: the average excess degree of the original graph ensemble must exceed 2. Recall that in terms of the moments of the degree distribution, $G_1^{\prime}(1) = (\langle k^2 \rangle - \langle k \rangle)/\langle k \rangle$, which permits to rewrite the criterion as $\langle k^2 \rangle > 3 \langle k \rangle$. As the case $L=0$ is symmetric under the substitution $\varphi \rightarrow 1-\varphi$, it is therefore not surprising that coexistence occurs whenever $\varphi_c^{(0)}<1/2$.
\begin{table*}
   \centering
   \caption{\label{tab:coex_databases}Description and properties of the databases used in Section~\ref{sec:coex_real} and in Figs.~\ref{fig:coex_real_nongeo} and \ref{fig:coex_real_geo}. The number of nodes ($N$), the average degree ($\langle k \rangle$), the highest degree ($k_\mathrm{max}$), the size of largest connected component ($S_\mathrm{max}$) as well as the value of $G_1^\prime(1)$ are given. The databases are divided into two categories: those whose behavior, with regards to observability, is closer to that of a random graph (top), and those whose behavior is similar to that of a lattice (bottom).}
   \begin{tabular}{l | r | c | r | r | r | c | c}
    \hline
    \hline
    \multicolumn{1}{c|}{Description} & \multicolumn{1}{c|}{$N$} & \multicolumn{1}{c|}{$\langle k \rangle$} & \multicolumn{1}{c|}{$k_\mathrm{max}$} & \multicolumn{1}{c|}{$G_1^\prime(1)$} & \multicolumn{1}{c|}{$S_\mathrm{max}$} & Fig. & Ref.  \\
    \hline
    Email communication network of Universitat Rovira i Virgili               &   1 133 & 9.08 &  1 080 &   125 &   1 133 & \ref{fig:coex_real_Email} & \cite{Guimera03_PhysRevE} \\
    Protein interaction network of \textit{S. cerevisiae}                     &   2 640 & 4.83 &    111 &  11.5 &   2 445 & \ref{fig:coex_real_ProteinCore} & \cite{Palla05_Nature} \\
    Web of trust of the Pretty Good Privacy (PGP) encryption algorithm        &  10 680 & 4.55 &    205 &  17.9 &  10 680 & --- & \cite{Boguna04_PhysRevE} \\
    Internet at the level of autonomous systems                               &  22 963 & 4.22 &  2 390 &   260 &  22 963 & \ref{fig:coex_real_Internet} & \footnote{Downloaded from \texttt{http://www-personal.umich.edu/\textasciitilde mejn/netdata/}.} \\
    arXiv co-authorship network                                               &  30 561 & 8.24 &    191 &  20.9 &  28 502 & \ref{fig:coex_real_arXiv} & \cite{Palla05_Nature} \\
    Gnutella peer-to-peer network                                             &  36 682 & 4.82 &     55 &  10.5 &  36 646 & --- & \cite{Ripeanu02_PeerToPeerSystems} \\
    Slashdot online social network                                            &  77 360 & 12.1 &  2 539 &   146 &  77 360 & \ref{fig:coex_real_Slashdot} & \cite{Leskovec09_InternetMathematics} \\
    Myspace online social network                                             & 100 000 & 16.8 & 59 108 &  3770 & 100 000 & --- & \cite{Ahn07_WWW07} \\
    Email exchange network from an undisclosed European research institution  & 265 009 & 2.75 &  7 636 &   536 & 224 832 & --- & \cite{Leskovec07_TKDD} \\
    World Wide Web                                                            & 325 729 & 6.69 & 10 721 &   280 & 325 729 & --- & \cite{Barabasi99_Science} \\
    &&&&&&\\
    Polish power grid                              &     3 374 & 2.41 & 11 & 2.15 &   3 374 & \ref{fig:coex_real_PolishGrid} & \cite{Zimmerman2011} \\
    Western States Power Grid of the United States &     4 941 & 2.67 & 19 & 2.87 &   4 941 & --- & \cite{Watts98_Nature} \\
    Road network of Pennsylvania                   & 1 088 092 & 2.83 &  9 & 2.20 & 1 087 562 & \ref{fig:coex_real_PennRoad} & \cite{Leskovec09_InternetMathematics} \\
    Road network of Texas                          & 1 379 917 & 2.79 & 12 & 2.15 & 1 351 137 & --- & \cite{Leskovec09_InternetMathematics} \\
    Road network of California                     & 1 965 206 & 2.82 & 12 & 2.17 & 1 957 027 & --- & \cite{Leskovec09_InternetMathematics} \\
    \hline 
    \hline
  \end{tabular}
\end{table*}
%
%
%
%
%
\subsection{Dependency on the depth $L$}
%
Using the results of Sec.~\ref{sec:coex_gen_math_from}, we now investigate the effect of varying $L$ on the coexistence regime. From Eqs.~\eqref{eq:coex_gen_epsilon}--\eqref{eq:coex_gen_chi}, it can be shown that for a fixed $\varphi$
\begin{align} \label{eq:coex_gen_decreasing_xi}
 \chi_i^{(L)} = \chi_i^{(L^\prime)} > \chi_{j}^{(L^\prime)}
\end{align}
for $0 \leq i \leq L+1$, $i < j \leq L^\prime+1$ and $L<L^\prime$. This implies that $g_i^{(L)}(\bm{x})=g_i^{(L^\prime)}(\bm{x})$ for $0 \leq i \leq L$ and $L<L^\prime$. Because $G_0(x)$ is a monotonous increasing function in [0,1] (as well as its derivatives), we directly see from Eqs.~\eqref{eq:coex_gen_w} that the fraction $w_{L+1}^{(L)}$ of non-occupied nodes decreases with increasing $L$
\begin{align*}
  \frac{ w_{L+1}^{(L)} }{ w_{L^\prime+1}^{(L^\prime)} }
    = \frac{ G_0\big(\chi_{L}^{(L)}\big) }{ G_0\big(\chi_{L^\prime}^{(L^\prime)}\big) }
    = \frac{ G_0\big(\chi_{L}^{(L^\prime)}\big) }{ G_0\big(\chi_{L^\prime}^{(L^\prime)}\big) } > 1
\end{align*}
for $L<L^\prime$. The more sparse non-occupied nodes are in the graphs, the more likely they will form finite-size components, therefore making an extensive component less likely. Hence we expect $\bar{\varphi}_c^{(L)}$ to decrease with increasing $L$. In fact, combining Eq.~\eqref{eq:coex_gen_gnc_threshold} and Eq.~\eqref{eq:coex_gen_decreasing_xi} leads to
\begin{align*}
 \frac{ 1-\bar{\varphi}_c^{(L)} }{ 1-\bar{\varphi}_c^{(L^\prime)} }
   = \frac{ G_1^\prime\big(\chi_{L^\prime}^{(L^\prime)}) }{ G_1^\prime\big(\chi_L^{(L)}) }
   = \frac{ G_1^\prime\big(\chi_{L^\prime}^{(L^\prime)}) }{ G_1^\prime\big(\chi_L^{(L^\prime)}) } < 1 \ ,
\end{align*}
which implies that $\bar{\varphi}_c^{(L)}>\bar{\varphi}_c^{(L^\prime)}$ for $L<L^\prime$. The emergence of the giant occupied component is affected in a similar way. As $L$ increases, directly and indirectly occupied nodes represent a larger fraction of the graphs (i.e., $1-w_{L+1}^{(L)}$) which increases the likelihood of an extensive component. We therefore expect $\varphi_c^{(L)} > \varphi_c^{(L^\prime)}$ for $L<L^\prime$. These insights are corroborated by Fig.~\ref{fig:coex_variousL}. Hence $\varphi_c^{(L)}$ is bounded from above by its value at $L=0$ [Eq.~\eqref{eq:coex_si_thresholds_L0}]. This is in accordance with the conclusion of Ref.~\cite{Yang12_PhysRevLett} where it is shown that $\varphi_c^{(1)}$ is bounded from above by a rapidly decreasing function of $G_1^\prime(1)$.

In order to assess the effect of varying the depth $L$ on the coexistence regime, we need to determine how $\varphi_c^{(L)}$ and $\bar{\varphi}_c^{(L)}$ behave relative to each other as $L$ increases. Unfortunately, although the Jacobian matrix determining the stability of the fixed point $\{a_{ij}\}=\bm{1}$ looks rather simple [cf. Eq.~\eqref{eq:coex_gen_goc_a}], we have not been able to completely settle this matter analytically. However, as illustrated by Fig.~\ref{fig:coex_threshold_vs_depth}, we find that, in all investigated scenarios, $\varphi_c^{(L)}$ decreases faster than $\bar{\varphi}_c^{(L)}$. If this behavior were to be proven true in general, it has the following implications. Firstly, if there is a coexistence interval for a given depth $L$, then there is a coexistence interval for all $L^\prime>L$. Consequently, increasing the depth $L$ cannot destroy the coexistence regime, it can only bring the bounds of its interval closer to $\varphi=0$. As a corollary, if $G_1^\prime(1)>2$, then there exists a coexistence interval for all depth. Note however that although $\varphi_c^{(L)}$ decreases faster than $\bar{\varphi}_c^{(L)}$, the width of the coexistence interval diminishes with increasing $L$ since both threshold values are decreasing [cf. Fig.~\ref{fig:coex_threshold_vs_depth}].

Secondly, if no coexistence interval exists for a given depth $L$, increasing the depth $L$ eventually creates a coexistence regime. This behavior is shown in Figs.~\ref{fig:coex_variousL_3}--\subref{fig:coex_variousL_5}. Thirdly, the symmetry of the case $L=0$ implies that both thresholds cannot be greater than 0.5 at the same time for any depth $L$ [see Fig.~\ref{fig:coex_threshold_vs_depth}]. As a final remark on the effect of the depth $L$ on the coexistence regime, the fact that $\varphi_c^{(L)}$ appears to be bounded from above by its value at $L=0$ implies that graphs whose degree distribution's second moment diverges (i.e., \textit{scale-free} degree distributions) would always have a coexistence interval. Indeed, whenever $G_1^\prime(1)\rightarrow\infty$, Eq.~\eqref{eq:coex_si_thresholds_L0} yields $\varphi_c^{(0)}=0$, and consequently $\varphi_c^{(L)}=0$ for any $L$. As \textit{heavy-tailed} degree distribution are ubiquitous in natural and technological complex systems \cite{Barabasi99_Science,Sornette04_CriticalPhenomena,Newman05_ContempPhys,Clauset09_SIAMRev,Hebert-Dufresne11_PhysRevLett,Hebert-Dufresne13a}, our analysis suggests that coexistence will be found in many real complex systems as $\varphi_c^{(L)}$ will be very close to zero for any depth $L$.
%
%
%
%
%
\section{Observability of real complex systems} \label{sec:coex_real}
%
To further our investigation, we simulated depth-1 percolation on graphs representing the underlying web of interactions of real complex systems. A total of 15 systems of diverse nature were considered; details of which are given in Table~\ref{tab:coex_databases}. Only a representative subset of our results on those systems are displayed in Figs.~\ref{fig:coex_real_nongeo}--\ref{fig:coex_real_geo}. Given a random sampling of a fraction $\varphi$ of the elements of a system (e.g., individuals, autonomous systems, proteins), the mathematical approach introduced in the previous sections allows one to estimate the \textit{coverage} of the system that is achieved given that information about the neighbors up to a distance $L$ of the sampled elements can be gathered as well. This coverage can be estimated in terms of the total number of elements about which information has been obtained (i.e., $\{w_i^{(L)}\}_{i=0,\ldots,L+1}$), or in terms of the largest number of contiguous elements (i.e., $S^{(L)}$), as in the main focus of this work.

Two examples will serve to explain the practical utility of our approach. Suppose that we want to get a global picture of the scientists working in a specific field without any prior information about that field. One way to achieve this is to browse the latest table of contents of appropriate journals, to identify scientists that have published something relevant to that field and then find with whom they co-authored papers during their careers. Although a sampling through the table of contents is not rigorously equivalent to the random and uncorrelated sampling considered in the previous sections, the quality of the coverage obtained can be estimated by studying depth-1 percolation on the associated co-authorship network. Looking up the co-authors of these co-authors up to a distance $L$ then corresponds to depth-$L$ percolation. Similarly, it has recently been revealed that intelligence agencies may gather information on individuals that are up to three ``hops'' (i.e., $L=3$) from suspected individuals \cite{Cho13_Science}. Again, our model offers a theoretical framework to estimate the extent of the population that could be investigated by studying the depth-$L$ percolation of online social networks, email communications or mobile phone networks.

Figures~\ref{fig:coex_real_nongeo}--\ref{fig:coex_real_geo} summarize the typical behaviors obtained when simulating depth-1 percolation on the graphs described in Table~\ref{tab:coex_databases}. Our results suggest that real systems behave differently with regards to observability according to whether they are geographically-constrained or not. Graphs that are not geographically-constrained behave more or less like random graphs (long-range connections are allowed), while geographically-constrained graphs behave more like lattices (no long-range connections).

We find that the observability of non-geographically-constrained graphs [Figs.~\ref{fig:coex_real_arXiv}--\subref{fig:coex_real_Slashdot}] is surprisingly well predicted by our mathematical framework despite the fact that most of these graphs have a far less trivial structure (e.g., clustering, correlations) than the Configuration Model which considers graphs that are random in all aspects other than the degree distribution. More importantly, we determine that these graphs have a structure that permits a coexistence regime. These graphs also display a vanishing threshold, $\varphi_c^{(1)}$, for the observable giant component. This agrees with the prediction of our model since these graphs have very skewed (i.e., scale-free) degree distribution [see Fig.~\ref{fig:coex_real_dd_powlaw}].

Contrariwise, our results for geographically-constrained graphs [Figs.~\ref{fig:coex_real_PennRoad}--\subref{fig:coex_real_PolishGrid}] display totally different behaviors \footnote{This conclusion could also be drawn by comparing the black curve of Fig.~2 with the solid gray curve of Fig.~3(a) in Ref.~\cite{Yang12_PhysRevLett}.}. Apart from the non-zero threshold for the occupied giant component, $\varphi_c^{(1)}$, caused by their approximatively exponential degree distributions [see Fig.~\ref{fig:coex_real_dd_expo}], the behavior of the two extensive components is poorly predicted by our mathematical framework. A fairly large coexistence interval is predicted while numerical simulations show that their structure does not, or barely, allow for a coexistence regime. Geographically-constrained graphs seem to be more accurately modeled by lattices than by random graphs. We have simulated depth-1 percolation on $L \times L$ square lattices where a fraction $p$ of edges are randomly removed. As shown in Fig.~\ref{fig:coex_real_lattice}, by simply choosing $p$ to match their average degree and $L$ to match their size, we have been able to qualitatively reproduce the results obtained with the real graphs [i.e., Fig.~\ref{fig:coex_real_PennRoad}--\subref{fig:coex_real_PolishGrid}]. Although preliminary, these results point towards the topological properties that should be incorporated in a future theoretical formalism to accurately model geographically-constrained graphs.
%
%
%
%
%
\section{Concluding remarks}
%
We have presented a general theoretical framework to study the observability of random graphs. On the one hand, it has allowed us to demonstrate that two extensive components, an observable and non-observable, may coexist for a wide range of realistic parameters, and that coexistence can be observed in many real complex systems. Our results suggest that coexistence could be an impediment to the monitoring of large real systems, and should therefore be considered in future investigations. On the other hand, the mapping of depth-$L$ percolation unto multitype graphs opens the way to the use of recent developments in percolation theory to study graphs with more realistic structures (e.g., clustering, correlations), and to investigate the efficiency of various distribution schemes for the monitoring units (e.g., according to the degree, to the local clustering or to the centrality of nodes) \cite{Allard12_JPhysA,Allard13b,Hebert-Dufresne13_PhysRevE,Hasegawa13_PhysRevE,Colomer-de-Simon13_SciRep,Serrano08_PhysRevLett}. We have also shown that our approach performs poorly at predicting the observability of geographically-constrained systems, and that achieving large-scale observability of these systems requires more monitoring units than suggested by calculations based on the Configuration Model. We have provided numerical evidences that these systems in fact behave more like lattices than random graphs. This observation raises many questions whose answers are expected to improve our understanding of the organization of these complex systems, and consequently to improve our capability to predict their behavior.
%
%
%
%
%
%
\begin{acknowledgments}
 The authors acknowledge the financial support of the Canadian Institutes of Health Research, the Natural Sciences and Engineering Research Council of Canada, and the Fonds de recherche du Qu\'ebec--Nature et technologies.
\end{acknowledgments}
%
%
%
%
%
\appendix
\section{Comparison with the approach of Yang \textit{et al.} for $L=2$} \label{sec:coex_YangVsAllard_L2}
%
In this section we analyze the solution to the observability problem given by Yang \textit{et al.} in the case $L=2$ and compare it with the prediction of our formalism. 
%
%
%
\subsection{Multitype formalism}
%
Let us first explicit the predictions of our approach. To lighten the presentation, we omit the superscript specifying the depth since this entire section focuses on the case $L=2$. Setting $L=2$ in Eq.~\eqref{eq:coex_gen_epsilon}, we obtain
\begin{subequations} \label{eq:coex_allard_L2_epsilon}
\begin{align}
  \varepsilon_{0} & = \varphi \\
  \varepsilon_{1} & = (1-\varphi) \big[ 1 - G_1(1-\varphi) \big]  \\
  \varepsilon_{2} & = (1-\varphi) \big[ G_1(1-\varphi) \nonumber \\
                  & - G_1\big((1-\varphi)G_1(1-\varphi)\big) \big] \ ,
\end{align}
\end{subequations}
where we have omitted the case $i=3$ since the present section focuses on the giant observable component solely. Similarly, Eq.~\eqref{eq:coex_gen_chi} becomes
\begin{subequations} \label{eq:coex_allard_L2_chi}
\begin{align}
  \chi_0 & = 1 \\
  \chi_1 & = 1 - \varphi \\
  \chi_2 & = (1-\varphi) G_1(1-\varphi) \\
  \chi_3 & = (1-\varphi) G_1\big((1-\varphi)G_1(1-\varphi)\big) \ ,
\end{align}
\end{subequations}
and Eq.~\eqref{eq:coex_gen_w} yields
\begin{subequations} \label{eq:coex_allard_L2_w}
\begin{align}
  w_{0} & = \varphi \\
  w_{1} & = (1-\varphi) \big[ 1 - G_0(1-\varphi) \big]  \\
  w_{2} & = (1-\varphi) \big[ G_0(1-\varphi) \nonumber \\
        & - G_0\big((1-\varphi)G_1(1-\varphi)\big) \big] \ .
\end{align}
\end{subequations}
Combining Eqs.~\eqref{eq:coex_allard_L2_epsilon}--\eqref{eq:coex_allard_L2_chi} with Eqs.~\eqref{eq:coex_gen_g} and \eqref{eq:coex_gen_goc_a}, we obtain the following system of equations
\begin{subequations} \label{eq:coex_allard_L2_a}
\begin{align}
  a_{00} & = G_1\big( \varphi a_{00} + (1-\varphi) a_{01} \big) \\
  a_{01} & = G_1(\varphi a_{00} + \varepsilon_1 a_{11} + \chi_2 a_{12}) \\
  a_{11} & = \frac{G_1(\varphi a_{00} + \varepsilon_1 a_{11} + \chi_2 a_{12}) - G_1(\varepsilon_1 a_{11} + \chi_2 a_{12})}{1 - G_1(1-\varphi)} \\
  a_{12} & = \frac{G_1(\varepsilon_1 a_{11} + \varepsilon_2 a_{22} + \chi_3)}{G_1(1-\varphi)} \\
  a_{22} & = \frac{G_1(\varepsilon_1 a_{11} + \varepsilon_2 a_{22} + \chi_3) - G_1(\varepsilon_2 a_{22} + \chi_3)}{G_1(1-\varphi) - G_1\big((1-\varphi)G_1(1-\varphi)\big)}
\end{align}
\end{subequations}
whose fixed point determines the size and behavior of the giant observable component. As for the case $L=1$, some $a_{ij}$ are equal: $a_{10}=a_{00}$ and $a_{21}=a_{11}$ when $L=2$. In fact, since the directly observable nodes (type 0) are randomly distributed and the type of the other nodes is inherited by the type of their neighbors, we find in general that $a_{i+1,i}=a_{i,i}$. In other words, the excess degree distribution of node A is independent of the type of the node from which it has been reached, as long as it is not the type of this neighbor that defines the type of node A. Note however that we will use interchangeably $a_{i+1,i}$ and $a_{i,i}$ according to whether it simplifies the notation or clarifies the significance of mathematical quantities. Combining Eqs.~\eqref{eq:coex_allard_L2_w} with Eqs.~\eqref{eq:coex_gen_g}--\eqref{eq:coex_gen_goc_S}, the size of the giant observable component is given by
\begin{align}
  S & = 1 - \varphi G_0\big( \varphi a_{00} + (1-\varphi) a_{01} \big) \nonumber \\
    &     - (1-\varphi) \Big\{ G_0(\varphi a_{00} + \varepsilon_1 a_{11} + \chi_2 a_{12}) \nonumber \\
    &     - G_0(\varepsilon_1 a_{11} + \chi_2 a_{12}) + G_0\big((1-\varphi)G_1(1-\varphi)\big) \nonumber \\
    &     + G_0(\varepsilon_1 a_{11} + \varepsilon_2 a_{22} + \chi_3) - G_0(\varepsilon_2 a_{22} + \chi_3) \Big\} \ .
\end{align}
%
%
%
\subsection{Yang \textit{et al.}'s approach}
%
Let us now recall the equations for $L=2$ as given in the Supplemental Material of Ref.~\cite{Yang12_PhysRevLett}. The authors define three probabilities $u$, $v$ and $s$ which are analogous to the $\{a_{ij}\}$ used in our approach: they correspond to the probability that a given randomly chosen edge does not lead to the giant observable component. These probabilities are defined as follows. (i) $u$ is the probability that an edge stemming from a node of type 0 (i.e., directly observable) does not lead to the giant observable component. (ii) $v$ is the probability that an edge stemming from a node of type 1 towards a node of type 1 or of type 2 does not lead to the giant observable component. (iii) $s$ is the probability that an edge stemming from a node of type 2 does not lead to the giant observable component. The authors then explain that by following a similar argument to the one used for $L=1$, it can be shown that
\begin{subequations}
\begin{align}
  u & = \varphi G_1(u) + (1-\varphi) G_1(\psi_1) \label{eq:coex_yang_L2_u} \\
  v & = G_1\big((1-\varphi)s\big) + \psi_2 \label{eq:coex_yang_L2_v} \\
  s & = G_1\big( (1-\varphi)G_1(1-\varphi) \big) + \psi_2 \nonumber \\
    & + G_1(\psi_3) - G_1\big( (1-\varphi)G_1(1-\varphi)s \big) \label{eq:coex_yang_L2_s}
\end{align}
where
\begin{align}
  \psi_1 & = \varphi G_1(u) + (1-\varphi)v \label{eq:coex_yang_L2_psi1} \\
  \psi_2 & = G_1(\psi_1) - G_1\big((1-\varphi)v\big) \label{eq:coex_yang_L2_psi2} \\
  \psi_3 & = (1-\varphi)\psi_2 + (1-\varphi)G_1(1-\varphi)s \label{eq:coex_yang_L2_psi3} \ ,
\end{align}
\end{subequations}
and that the size of the giant observable component is given by
\begin{align}
  S_Y & = 1 - \varphi G_0(u) 
        - (1-\varphi) \Big\{ G_0(\psi_1) - G_0\big((1-\varphi)v)\big) \nonumber \\
      & + G_0(\psi_3) - G_0\big( (1-\varphi)G_1(1-\varphi)s \big) \Big\} \nonumber \\
      & + G_0\big((1-\varphi)G_1(1-\varphi)\big) \ .
\end{align}
%
%
%
\subsection{Comparison of the two approaches}
%
We now investigate whether these two approaches are equivalent or not. As mentioned above, $u$ is the probability that a directly observable node (type 0) is not linked to the giant observable component via one specific edge. This corresponds to the probability that the node at the other end of the edge, say node B, is of type 0 (probability $\varepsilon_0$) and that the edge does not lead to the giant observable component (probability $a_{00}$), or that node B is of type 1 (probability $\chi_1$) and that the edge does not lead to the giant observable component (probability $a_{01}$). Summing these two contributions and then using Eqs.~\eqref{eq:coex_allard_L2_a}, we find
\begin{align}
  u & = \varepsilon_0 a_{00} + \chi_1 a_{01} \nonumber \\
    & = \varphi G_1\big( \underbrace{\varphi a_{00} + (1-\varphi) a_{01}}_{u} \big) \nonumber \\
    & + (1-\varphi) G_1( \underbrace{\varphi a_{10} + \varepsilon_1 a_{11} + \chi_2 a_{12}}_{\psi_1} ) \ ,
\end{align}
which corresponds to Eq.~\eqref{eq:coex_yang_L2_u} provided that the identification of $\psi_1$ holds. As in the case $L=1$, $\psi_1$ is the probability that a node of type 1 is not connected to the giant observable component via a specific edge. Three different scenarios must be accounted for depending on the type of the node at the other end of the edge: this node can be of type 0, type 1 or type 2, with probability $\varepsilon_0$, $\varepsilon_1$ and $\chi_2$, respectively. Multiplying each probability by the corresponding probability that the edge does not lead to the giant component, we retrieve the above identification
\begin{align}
  \psi_1 & = \varepsilon_0 a_{10} + \varepsilon_1 a_{11} + \chi_2 a_{12} \nonumber \\
         & = \underbrace{\varphi G_1\big( \varphi a_{00} + (1-\varphi) a_{01} \big)}_{\varphi G_1(u)} + \underbrace{\varepsilon_1 a_{11} + \chi_2 a_{12}}_{(1-\varphi)v} \ ,
\end{align}
which corresponds to Eq.~\eqref{eq:coex_yang_L2_psi1} provided that the identification of $v$ holds. The first term on the right-hand side of this last equation corresponds to the situation where the node at the other end of the edge is of type 0 (probability $\varphi$) and does not lead to the giant observable component [probability $G_1(u)$]. The second term corresponds to the case where the neighboring node is of type 1 or of type 2, which occurs with probability $1-\varphi$ (recall that the neighbor of a node of type 1 cannot be of type 3, by definition), and that this edge does not lead to the giant component, which by definition occurs with probability $v$. In terms of the formalism that we propose, the probability for an edge leaving a node of type 1 to lead to a node of type 1 is $\varepsilon_1$, and is $\chi_2$ if the neighboring node is of type 2 instead. Weighting these probabilities with the appropriate probability that the edge does not lead to the giant observable component yields precisely
\begin{widetext}
\begin{align} \label{eq:coex_correspondence_L2_v}
  (1-\varphi) v & = \varepsilon_1 a_{11} + \chi_2 a_{12} \nonumber \\
                & = (1-\varphi) \Big\{ \underbrace{\underbrace{G_1(\varphi a_{10} + \varepsilon_1 a_{11} + \chi_2 a_{12})}_{G_1(\psi_1)} - G_1(\underbrace{\varepsilon_1 a_{11} + \chi_2 a_{12}}_{(1-\varphi)v})}_{\psi_2} + G_1(\underbrace{\varepsilon_1 a_{21} + \varepsilon_2 a_{22} + \chi_3}_{(1-\varphi)s} )\Big\} \ ,
\end{align}
from which we retrieve Eq.~\eqref{eq:coex_yang_L2_psi2} and Eq.~\eqref{eq:coex_yang_L2_v} provided that the identification of $s$ holds. We see from this last equation that $\psi_2$ is the probability that a node of type 1 reached from a node of type 1 does not lead to the giant observable component. Note that because it has been reached from a node of type 1, this node must have at least one neighbor of type 0 in order to be of type 1. Since $G_1(\psi_1)$ includes the case where all neighbors of a node of type 1 are not of type 0, the probability of such an event must be removed from the count, which is achieved by subtracting $G_1\big((1-\varphi)v\big)$. Additionally if the node at the other end of the edge is of type 2 instead of type 1, then none of its \textit{other} neighbors must be of type 0, which occurs individually with probability $1-\varphi$, and must not lead to the giant component, which by definition occurs with probability $s$. Averaging over the number of \textit{other} neighbors [the excess degree distribution generated by $G_1(x)$], we obtain the third term on the right-hand side of Eq.~\eqref{eq:coex_correspondence_L2_v}. Again, the probability $(1-\varphi)s$ can be expressed in terms of our formalism. The probability that an edge leaving a node of type 2 towards a node of type 1, of type 2 and of type 3 is respectively $\varepsilon_1$, $\varepsilon_2$ and $\chi_3$. Weighting $\varepsilon_1$ and $\varepsilon_2$ by the probability that the edge does not lead to the giant component (recall that a node of type 3 does not belong to the giant observable component ``with probability 1'') yields our previous identification
\begin{align} \label{eq:coex_correspondance_L2_s}
  (1-\varphi)s & = \varepsilon_1 a_{21} + \varepsilon_2 a_{22} + \chi_3 \nonumber \\
               & = (1-\varphi) \Big\{ \psi_2 + \big[G_1\big(\underbrace{(1-\varphi)\psi_2 + \varepsilon_2 a_{22} + \chi_3}_{\psi_3}\big) - G_1(\varepsilon_2 a_{22} + \chi_3)\big] + G_1\big((1-\varphi)G_1(1-\varphi)\big) \Big\} \ ,
\end{align}
\end{widetext}
where we have used the fact that $\varepsilon_1 a_{21}=\varepsilon_1 a_{11}=(1-\varphi)\psi_2$ and the definition of $\chi_3$. Comparing this last equation with Eqs.~\eqref{eq:coex_yang_L2_s} and \eqref{eq:coex_yang_L2_psi3}, we find that the two approaches are equivalent if
\begin{align} \label{eq:coex_corresondance_L2_missing_link}
  (1-\varphi)G_1(1-\varphi)s = \varepsilon_2 a_{22} + \chi_3 \ .
\end{align}
As for $\psi_2$, we see from Eq.~\eqref{eq:coex_correspondance_L2_s} that $\psi_3$ is the probability that an edge between two nodes of type 2 does not lead to the giant observable component. Since the node of type 2 reached from such edge does not inherit its type from the node of type 2 at the other end of the edge, at least one of its \textit{other} neighbors must be of type 1. Again, since $G_1(\psi_3)$ includes the configuration where every \textit{other} neighbors of the node of type 2 are of type 2 or of type 3, this eventuality must be removed from the count, which is achieved by subtracting $G_1(\varepsilon_2 a_{22} + \chi_3)$.

Let us now investigate the validity of Eq.~\eqref{eq:coex_corresondance_L2_missing_link}. Replacing $(1-\varphi)s$ by $\varepsilon_1 a_{21} + \varepsilon_2 a_{22} + \chi_3$ yields the following alternative criterion for the complete equivalence of the two approaches
\begin{align}
  \varepsilon_2 a_{22} + \chi_3 = (\varepsilon_1 a_{21} + \varepsilon_2 a_{22} + \chi_3) G_1(1-\varphi) \ ,
\end{align}
which is most certainly not true in general. Although very similar, these two approaches are therefore not strictly equivalent. In fact, their numerical predictions differ by less than a fraction of one percent in all investigated cases. This difference stems for the use of $s$ for two different purposes in the approach presented in Ref.~\cite{Yang12_PhysRevLett}. On the one hand, $s$ is initially defined as the probability that an edge stemming out of a node of type 2 does not lead to the giant observable component irrespective of the type of the node at its other end (i.e., type 1, type 2 or type 3). On the other hand, as it is used in Eqs.~\eqref{eq:coex_yang_L2_s} and \eqref{eq:coex_yang_L2_psi3}, the possibility that the type of the node at the other end is of type 1 is excluded since it is taken care of by the probability $(1-\varphi)\psi_2$. More precisely, in our formalism $(1-\varphi)G_1(1-\varphi) = \varepsilon_2 + \chi_3$ is the probability that the node at the other end of an edge and its \textit{other} neighbors are not of type 0. Since this edge is leaving a node of type 2, the node at its other end is of type 2 or of type 3; it therefore cannot be of type 1. In other words, the term $(1-\varphi)G_1(1-\varphi)s$ uses $s$ as an approximation of the probability that an edge leaving a node of type 2 towards a node of type 2 or type 3 does not lead to the giant observable component.
%
%
%
%
%
%
%
%
%
%
\end{document}